\documentclass[reprint,amsmath,nofootinbib,amssymb,aps,showpacs,prc]{revtex4-1}
\usepackage{graphicx}
\usepackage{subfigure}
\usepackage{multirow}
\usepackage{mathrsfs}
\usepackage[colorlinks,
linkcolor=blue,
anchorcolor=blue,
citecolor=blue]{hyperref}
\usepackage[all]{hypcap}
\usepackage{epstopdf}
\usepackage{caption}
\usepackage{color}
\usepackage{float}
\usepackage{bm}

\begin{document}


\title{Probe Chiral Magnetic Effect with Signed Balance Function}

\author{A. H. Tang}
\affiliation{
\mbox{Brookhaven National Laboratory, Upton, New York 11973, USA}\\
}

\begin{abstract}
In this paper a pair of observables are proposed as alternative ways, by examining the fluctuation of net momentum-ordering of charged pairs, to study the charge separation induced by the Chiral Magnetic Effect (CME)  in relativistic heavy ion collisions. They are, the out-of-plane to in-plane ratio of fluctuation of the difference between signed balance functions measured in pair’s rest frame, and the ratio of it to similar measurement made in the laboratory frame. Both observables have been studied with simulations including flow-related backgrounds, and for the first time, backgrounds that are related to resonance's global spin alignment. The two observables have similar positive responses to signal, and opposite, limited responses to identifiable backgrounds arising from resonance flow and spin alignment. Both observables have also been tested with two realistic models, namely, a  multi-phase  transport (AMPT) model and the anomalous-viscous fluid dynamics (AVFD) model. These two observables, when cross examined, will provide useful insights in the study of CME-induced charge separation.
\end{abstract}


\maketitle

\section{Introduction}\label{sec:intro}

It has been pointed out that the hot and dense matter created in relativistic heavy-ion collisions may form metastable domains where the parity and
time-reversal symmetries are locally violated~\cite{Kharzeev:1998kz}, creating fluctuating, finite topological charges. In noncentral collisions, when such domains interplay with the ultra-strong magnetic fields produced by spectator protons~\cite{Kharzeev:2007jp}, they
can induce electric charge separation parallel to the magnetic field direction --- the chiral magnetic effect (CME)~\cite{Kharzeev:2007jp,Kharzeev:2015znc,Huang}. 

In an event, charge separation along the magnetic field direction may be described by sine terms in the Fourier decomposition of the charged-particle azimuthal distribution
\begin{eqnarray}
\begin{aligned}
\frac{dN_{\pm}}{d\Delta \phi} \propto  & 1  + 2v_{1}\mathrm{cos}(\Delta \phi) + 2v_{2}\mathrm{cos}(2\Delta \phi) \\ 
&+ 2v_{3}\mathrm{cos}(3\Delta \phi)  
 + \cdots + 2 a_{\pm}\mathrm{sin}(\Delta \phi) + \cdots,
\end{aligned}
\label{eq:a1}
\end{eqnarray}
where $\Delta \phi$ ($ = \phi - \Psi_{RP}$) is particle's azimuthal angle with respect to the reaction plane ($\Psi_{RP}$). The reaction plane is defined by the beam direction and the line connecting the centroids of two colliding nuclei at their closest approach (impact parameter $\hat{b}$). The direction of magnetic field $B$ is perpendicular to the reaction plane. $v_{1}$, $v_{2}$ and $v_{3}$ are
coefficients accounting for the directed, elliptic and triangular flow~\cite{Poskanzer:1998yz}, respectively. The $a_{\pm}$ ($a_{+} = - a_{-}$) parameter describes the charge separation effect. In a parity violating domain, net-positive and net-negative topological charges can be produced with equal likelihood, causing the sign of $a_{\pm}$ to fluctuate from event to event depending on event's net topological charge. This makes $a_{\pm}$ not distinguishable on an event-by-event basis. However one can instead study the effect of $a_{\pm}$'s fluctuation $\langle a_{1}^2 \rangle$, where $a_{1} \equiv |a_{+}| \equiv |a_{-}|$, noting that $\langle a_{+} \rangle = \langle a_{-} \rangle = 0$.

To study the CME experimentally one has to look for the enhanced fluctuation of charge separation in the direction perpendicular to the reaction plane, relative to the fluctuation in the reaction plane itself. This is the basis of all CME searches in heavy-ion collisions, including the method under discussion in this paper. Experimental searches for the CME have been going on for a decade, with multiple methods\cite{Voloshin:2004vk, Adamczyk:2013hsi,Li:2018oot,Xu:2017qfs,Adamczyk:2013kcb,Magdy:2017yje} and carried out by experiments at both RHIC\cite{Abelev:2009ac,Abelev:2009ad,Adamczyk:2013hsi,Adamczyk:2013kcb,Adamczyk:2014mzf} and LHC\cite{Abelev:2012pa,Sirunyan:2017quh,Khachatryan:2016got}. So far there is no conclusive evidence for the existence of the CME in heavy ion collisions, see\cite{Kharzeev:2015znc} for a progress review. The major challenge in the CME searches is that backgrounds, in particular those related to elliptic flow of resonances, can produce similar enhancement in fluctuation in the direction perpendicular to the reaction plane\cite{Bzdak:2012ia,Pratt:2010zn,Schlichting:2010qia,Wang:2009kd,Wang:2016iov,Feng:2018chm}. To have the background under control, the STAR experiment at RHIC has collected collisions from isobaric collisions and the data analysis is on-going.

\begin{figure}[htbp]
\centering
\makebox[1cm]{\includegraphics[width=0.2 \textwidth]{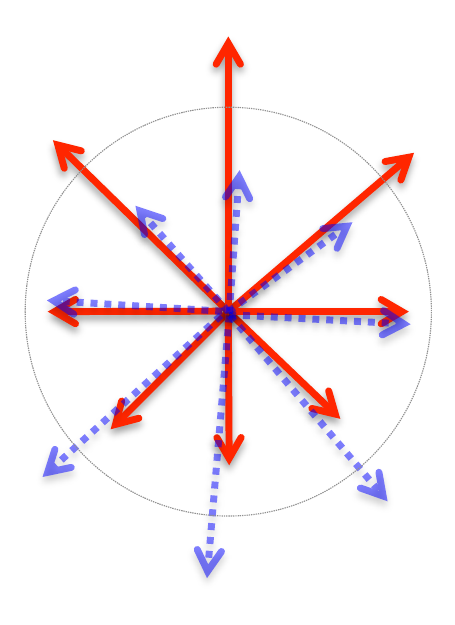}}
\caption{(Color online) Cartoon illustration of an event with the CME but particles are emitted isotropically. Solid (dashed) lines represent  momentum of positive (negative) particles. The circle is to guide eyes for the illustration of an isotropic source in azimuth.}
\label{fig:ChargeSeparation_IsotropicEmission-crop}
\end{figure}

For a pair of positive and negative particles originated from parity-odd domain, the CME increases the momentum of one of  them in $+\hat{B}$ direction, and the other, in the $-\hat{B}$ direction, causing a separation. Here the phenomenon in focus is the separation in momentum. However, all previous CME searches are based on azimuthal angle correlations, not directly on the separation in momentum. This is so as long as one writes down in the first place the $\frac{dN_{\pm}}{d\Delta \phi}$ distribution in the form of Eq. (\ref{eq:a1}). Finite $a_{\pm}$ in Eq. (\ref{eq:a1}) characterizes a pattern where positive particles and negative particles diverge in opposite directions along $B$, similar to directed flow but in the out-of-plane direction. By characterizing the CME with a flow-like pattern, one takes advantage of vast, existing knowledge on flow which is sophisticated and convenient to manipulate with. That said, there are also limitations associated with these approaches. This can be illustrated with an example below. Fig.~\ref{fig:ChargeSeparation_IsotropicEmission-crop} shows an event in which particles are emitted isotropically in azimuth. The CME causes all positive (negative) particles to have an extra increase in momentum upwards (downwards). However, if the only available information is the azimuthal angle, it is not possible to identify this as a case of charge separation. This problem can be alleviated by having extra restrictions on particle's momentum, but it is not as tempting as considering the separation in momentum space directly, which is the focus of this study.

In this paper, a pair of observables are proposed to study the CME effect. One of them is
the out-of-plane to in-plane ratio of fluctuation of the difference between signed balance functions measured in the particle pair's rest frame, and the other is the ratio of it to the similar measurement made in the laboratory frame. Here the sign of a balancing pair is determined by their momentum ordering in the direction perpendicular to the reaction plane, and the difference between signed balance functions accounts for net momentum-ordering. The two observables are shown to have positive responses to signal, but opposite, limited responses to identifiable backgrounds arising from resonance flow and global spin alignment. In following sections the signed balance function will be described first and the two observables will be introduced, followed by the discussion of toy model studies with various background scenarios, including flow related backgrounds and a background that is caused by resonance's global spin alignment. The latter has not been considered previously. Results from realistic models will be presented, including one with pure background, namely a multi-phase transport (AMPT) model~\cite{Lin:2004en}, as well as another with both signal and background, namely the anomalous-viscous fluid dynamics (AVFD) model ~\cite{Jiang:2016wve,Shi:2017cpu}. The merits of the observables as well as their limitations are summarized at the end.

\section{Signed balance function}\label{sec:balanceFunction}

The balance function (BF), in its general form, describes the absolute separation of particles in phase space~\cite{Bass:2000az,Adams:2003kg}. At RHIC and LHC, the balance function in pseudorapidity, $B(\Delta \eta)$, which spans the absolute difference in pseudorapidity between two balancing particles, $\Delta \eta = |\eta_a - \eta_b|$, is usually used to study the delayed hadronization in head-on collisions\cite{Adams:2003kg,Abelev:2010ab, Aggarwal:2010ya,Adamczyk:2015yga,Abelev:2013csa}. The signed balance function, previously proposed to study the magnetic field in heavy ion collisions\cite{Ye:2018jwq}, considers the signed difference instead of the absolute separation of particles in phase space. 

Before going further in details, let's first introduce the coordinate system used in this paper. The $x$-axis is set by the direction of the impact parameter ($\hat{b}$) which is also the direction of the reaction plane. The $z$-axis represents the beam direction ($\hat{p}_{\mathrm{beam}}$), and the $y$-axis ($\hat{y}= - \hat{b} \times \hat{p}_\mathrm{beam}$) is perpendicular to the reaction plane. The magnetic field direction, as well as the global angular momentum vector, are pointing in ($-\hat{y}$) direction. With this setup, the charge separation due to the CME is along the $y$-axis. This setup is the same as in \cite{Ye:2018jwq}. 

The signed balance function approach consists of four steps:

\textbf{1) Count pair's momentum-ordering. } Considering the momentum in $y$ direction ($p_y$) for now, for any two particles $\alpha$ and $\beta$, if $p^\alpha_y >  p^\beta_y$, then $\alpha$ is considered leading $\beta$, otherwise, tailing it. Two signed balance functions can be invoked to ``count" the pair's momentum-ordering,

\begin{eqnarray}
\begin{aligned}
B_{P} (S) =  \frac{N_{+-}(S)-N_{++}(S)}{N_+},
\end{aligned}
\label{eq:Bp}
\end{eqnarray}
and
\begin{eqnarray}
\begin{aligned}
B_{N} (S) =  \frac{N_{-+}(S)-N_{--}(S)}{N_-}.
\end{aligned}
\label{eq:Bn}
\end{eqnarray}
Here the subscript $P$ and $N$ stand for positive and negative terms, respectively.
For a given term $N_{\alpha\beta}$, $S=+1$ if $\alpha$ is leading $\beta$ and $S=-1$ if it is tailing it.  $N_{+(-)}$ is the number of positive (negative) particles in an event. 

\textbf{2) Count net momentum-ordering for each event. } An event by event difference between $B_{P}$ and $B_{N}$ can be calculated :

\begin{eqnarray}
\begin{aligned}
\delta B(\pm 1) =  B_{P}(\pm 1)-B_{N}(\pm 1),
\end{aligned}
\label{eq:deltaB_pm}
\end{eqnarray}
and
\begin{eqnarray}
\begin{aligned}
\Delta B =  \delta B(+1) - \delta B(-1).
\end{aligned}
\label{eq:deltaB}
\end{eqnarray}

\begin{figure}[htbp]
\centering
\makebox[1cm]{\includegraphics[width=0.45 \textwidth]{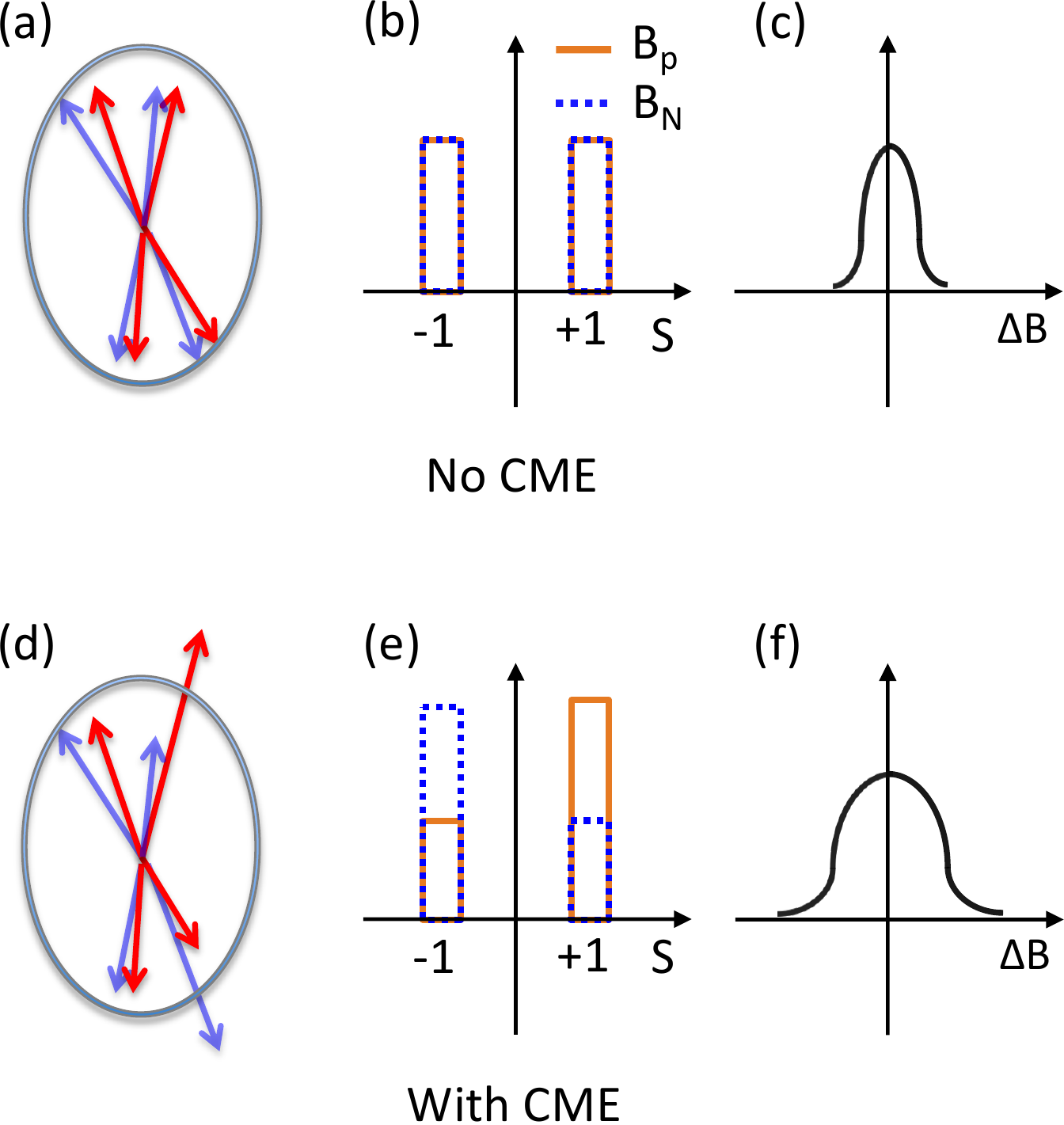}}
\caption{(Color online) Cartoon illustration of positive (red) and negative (blue) particle directions (left panels (a) and (d)) and $S$ distributions (middle panels (b) and (e)) of an event, as well as the $\Delta B$ distribution over many events (right panels (c) and (f)). The top row is for the case without the CME, and the bottom one, with it.  }
\label{fig:BF_cartoon}
\end{figure}
$\Delta B$ accounts for the normalized net momentum-ordering for each event. When there is no CME effect, for a positive-negative particle pair, the probability of the positive particle leading the negative one equals the probability of tailing it. This means that $B_P$ and $B_N$ are in principle measuring the same quantity, and the distribution of $\Delta B$ is only subject to statistical fluctuation (top row of Fig.~\ref{fig:BF_cartoon}). When there is CME effect, within an event the two probabilities become unbalanced, resulting more pairs with particles of one charge-type leading than tailing the other type. This makes for each event $B_P$ and $B_N$ to differ from each other, and as a consequence, the distribution of $\Delta B$ has a broadened width (bottom row of Fig.~\ref{fig:BF_cartoon}). 

\begin{figure}[htbp]
\centering
\makebox[1cm]{\includegraphics[width=0.45 \textwidth]{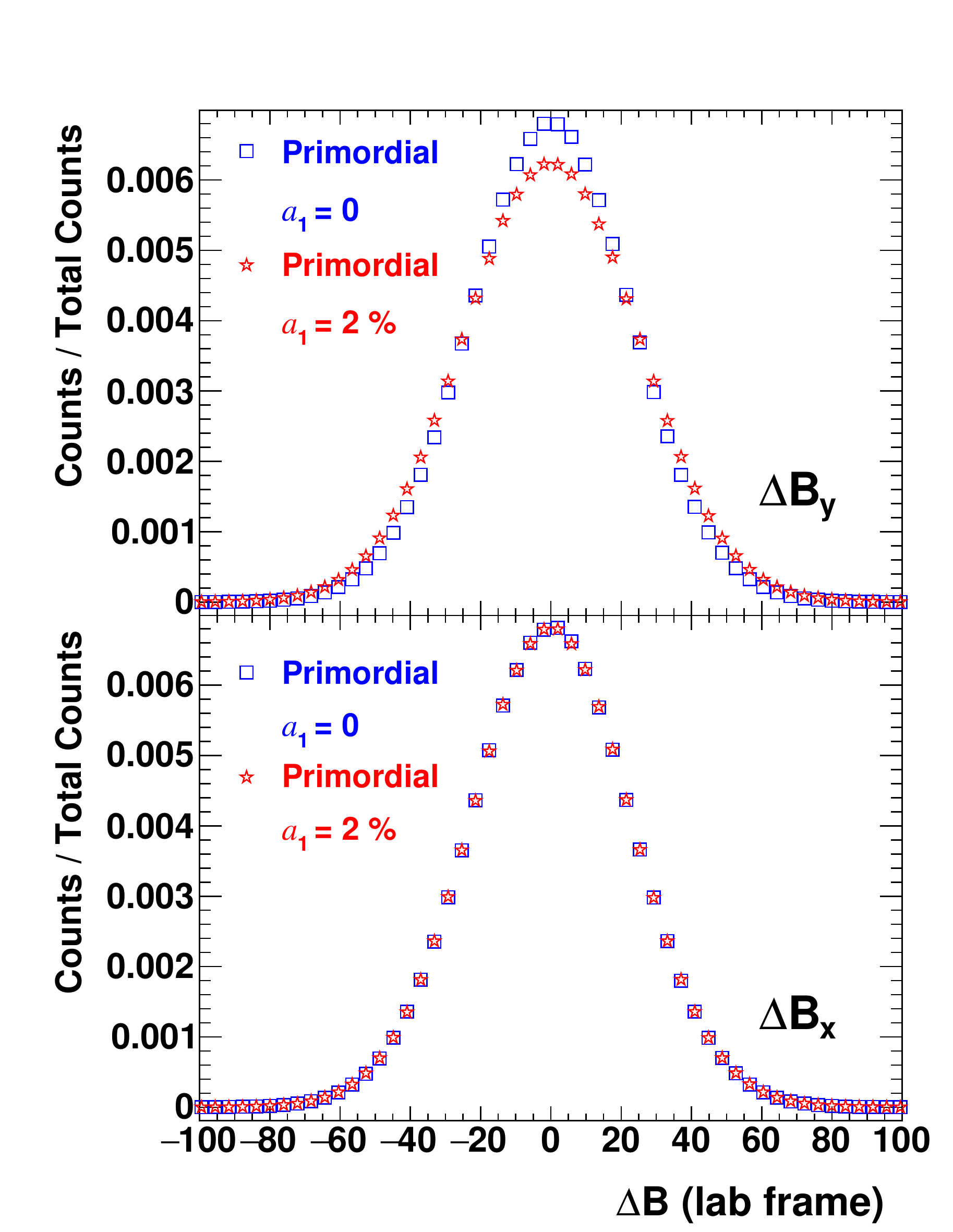}}
\caption{(Color online) $\Delta B_x$ (bottom) and $\Delta B_y$ (top) distributions for two cases of simulated events, one without the CME ($a_1 = 0$), and the other, with the CME ($a_1 = 2\%$). }
\label{fig:BFHisto_lab_example}
\end{figure}

The distribution of $\Delta B$ can be obtained for both $x$ ($\Delta B_{x}$) and $y$ direction ($\Delta B_{y}$). The distribution of $\Delta B_{x}$ is not broadened as there is no charge separation in the $x$ direction. Fig.~\ref{fig:BFHisto_lab_example} shows the distribution of $\Delta B_{x}$ and $\Delta B_{y}$ with simulated events. One can see that when a finite $a_1$ is applied to primordial particles, the distribution of $\Delta B_y$ is broadened relative to $\Delta B_x$. The setup for the simulation is deferred to section~\ref{sec:ToyModelSimulation} where it is described together with other simulation setups.

\textbf{3) Look for enhanced event-by-event fluctuation of net momentum-ordering in $y$ direction. } To cancel out the statistical fluctuation, one can calculate the ratio of the width of the distribution of $\Delta B_{y}$ to that of $\Delta B_{x}$,
\begin{eqnarray}
\begin{aligned}
r= \sigma_{\Delta B_y} / \sigma_{\Delta B_{x}}.
\end{aligned}
\label{eq:r}
\end{eqnarray}

 $r$ will be unity for the case without the CME, and greater than unity for the case with it. The strength of the CME will be positively correlated with $r$'s deviation from unity.
\begin{figure}[htbp]
\centering
\makebox[1cm]{\includegraphics[width=0.45 \textwidth]{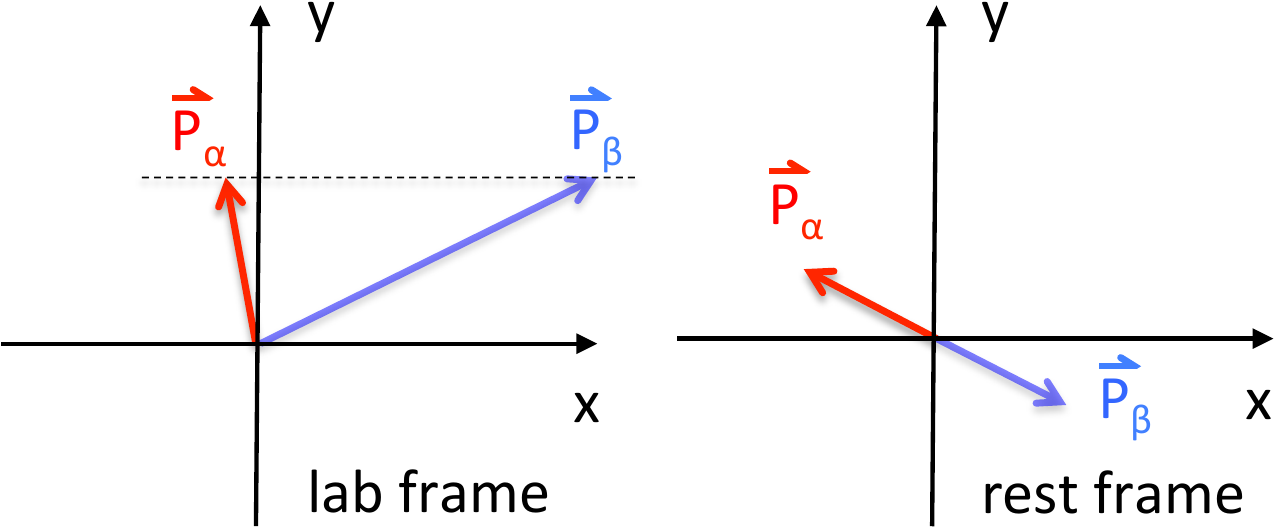}}
\caption{(Color online) Cartoon illustration of a pair viewed in the laboratory frame (left) and pair's rest frame (right). }
\label{fig:boost_cartoon}
\end{figure}

\textbf{4) Compare results obtained with different frames. } The ratio $r$ can be calculated in the laboratory frame ($r_{\mathrm{lab}}$) and pair's rest frame ($r_{\mathrm{rest}}$). Here for the signed balance function approach, it is argued that the rest frame is the most appropriate frame to study charge separations. This can be understood in an intuitive way -- the clearest observation of two particles moving away from each other has to be, naturally, made by an observer who is at rest with the two-particle system under consideration. Boosting from rest frame to lab frame does not always preserve the correct ordering in $P_y$. In Fig.~\ref{fig:boost_cartoon} an example is given to illustrate this point. The cartoon on the left depicts a pair in the laboratory frame, and it is not counted as a case of charge separation by the signed BF approach, as both particles have the same $p_y$. When the same pair is viewed in the rest frame (right cartoon), it is clearly a case of charge separation. Note in the rest frame two particles are traveling back-to-back, and in this particular frame leading(tailing) simply means particle traveling in positive(negative) $\hat{y}$ direction -- making it easy to be identified in the signed BF approach. Indeed by definition $r_{\mathrm{rest}}$ is always the most sensitive one when responding to real charge separation, however, it is not guaranteed so when responding to backgrounds --  $r_{\mathrm{rest}}$ may lag behind $r_{\mathrm{lab}}$. This will be discussed in details later in this paper. 

\begin{figure}[htbp]
\centering
\makebox[1cm]{\includegraphics[width=0.45 \textwidth]{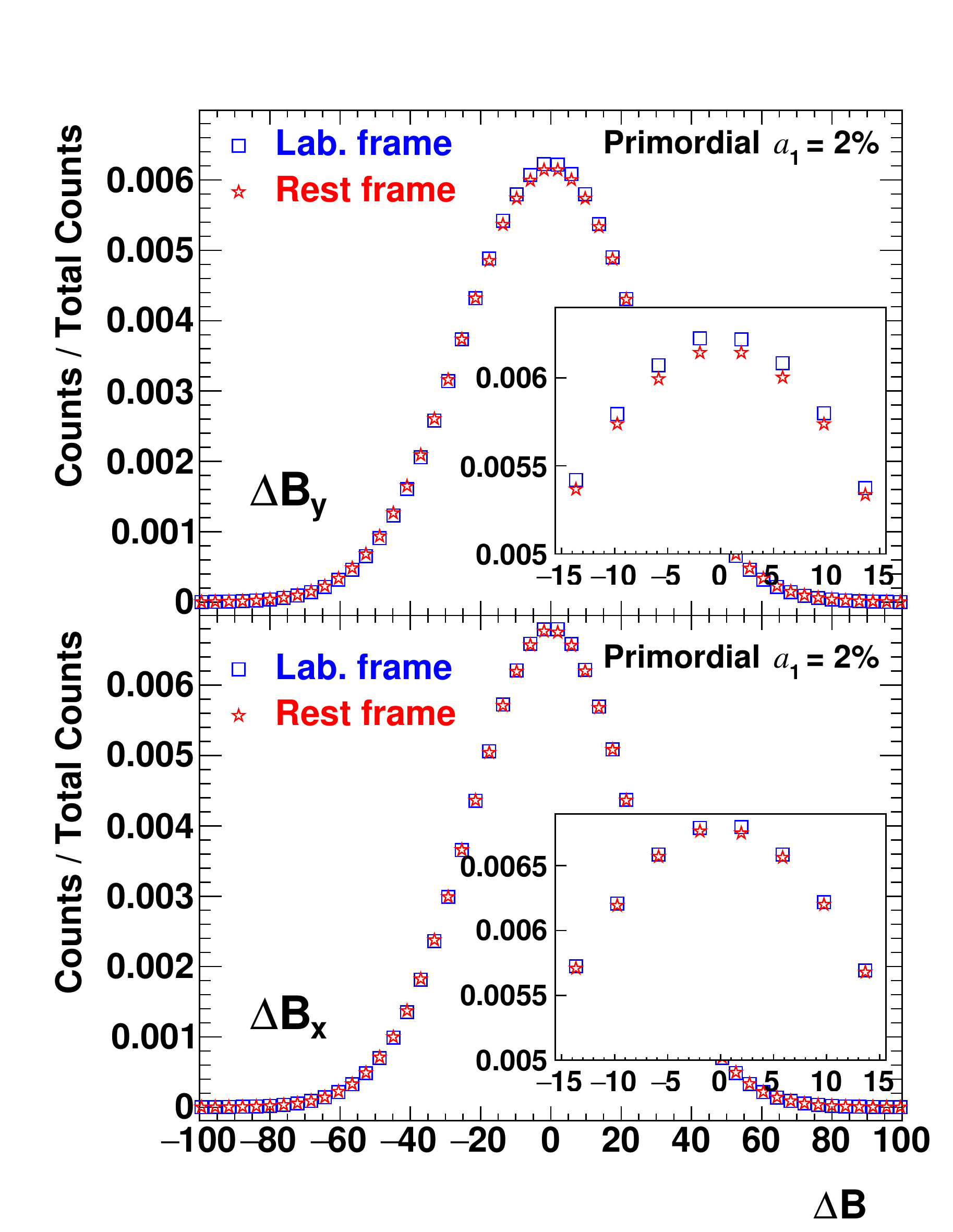}}
\caption{(Color online) $\Delta B_x$ (bottom) and $\Delta B_y$ (top) distribution obtained in laboratory frame (blue) and pair's rest frame (red), for simulated events with finite CME ($a_1 = 2\%$). Insets are magnified views of peak regions. }
\label{fig:BFHisto_rest_lab_example}
\end{figure}

Fig.~\ref{fig:BFHisto_rest_lab_example} shows $\Delta B_x$ and $\Delta B_y$ distributions obtained in laboratory and pair's rest frame, for simulated events with finite CME effect. It can be seen that $\Delta B_y$ distribution in rest frame is broadened  more than that in laboratory frame, although such effect is relatively small if compared to the difference in width between $\Delta B_x$ and $\Delta B_y$ in either frame.

It would be useful to calculate the ratio of the two,
\begin{eqnarray}
\begin{aligned}
R_{B} \equiv  \frac{r_{\mathrm{rest}}}{r_{\mathrm{lab}}},
\end{aligned}
\label{eq:R_B}
\end{eqnarray}
where the subscript ``B" stands for Balance Function.
It will be shown below with simulations that while $R_{B}$
responds positively to signal (like each of $r_{\mathrm{rest}}$ and $r_{\mathrm{lab}}$ themselves does), it responds in the opposite direction (relative to $r_{\mathrm{rest}}$  and $r_{\mathrm{lab}}$) to backgrounds arising from resonance flow and global spin alignment. This information can be useful under certain scenarios in identifying charge separation induced by backgrounds. For example, if $r_{\mathrm{rest}}$ is above unity and $R_{B}$ is below it (or vice versa), then it is an indication of background contribution. On the other hand, if both $r_{\mathrm{rest}}$ and $R_{B}$ are above unity, then we have a case in favor of the CME.
 
For convenience, in this paper and at a few places, either of the three ratios being above unity, which can be caused by the CME and/or background, will be referred to as apparent charge separation. The apparent charge separation is what is usually measured in experiments.

 \section{Toy model simulations}\label{sec:ToyModelSimulation}
 
 In this section, a series of toy model simulations for various signal and/or background scenarios will be presented. The section starts with simple cases followed by cases with relatively more realistic considerations. To avoid making conclusions by accident, in each case observables are studied against the change of only one parameter while everything else is unchanged. For all cases, a simulated event consists of 324 primordial charged pions (162 for each charge type), and 33 $\rho$ resonances, each of which decays into a $\pi^{+} + \pi^{-}$ pair. This configuration gives a total multiplicity that matches the multiplicity within 2 units of rapidity for $30-40\%$ central Au+Au collisions at $\sqrt{s_{NN}} = 200 $ GeV ~\cite{Abelev:2008ab}, while maintaining the ratio in yield of $\rho$ resonance to negative particles at $\sim17\%$~\cite{Adams:2003cc}. The decay of $\rho \rightarrow \pi^{+} + \pi^{-}$ is implemented with PYTHIA6~\cite{Sjostrand:2006za}. Primordial pions and $\rho$ resonances are allowed to have their own $v_{2}$ and $v_{3}$, and in addition, primordial pions can have finite CME signal ($a_{1}>0$), and $\rho$ resonances can have finite global spin alignment ($\rho_{00} \neq 1/3$)~\cite{Liang:2004ph,Liang:2004xn,Liang:2007ma,Betz:2007kg,Gao:2007bc,Becattini:2013vja}. Unless otherwise specified, following ~\cite{Wang:2016iov}, primordial pions are generated according to a Bose-Einstein distribution~\cite{Abelev:2008ab}, $dN_{\pi^{\pm}}/dm_{T}^2 \propto (e^{m_{T}/T_{BE}} -1)^{-1}$, where $m_{T}=\sqrt{p_{T}^2+m_{\pi}^2}$ ($m_{\pi}$ is the $\pi^{\pm}$ rest mass), and $T_{BE}$ is set to be 212 MeV in order to have a $\langle p_{T} \rangle$ of 400 MeV~\cite{Abelev:2008ab}. $\rho$ resonances are generated according to $dN_{\rho}/dm_{T}^2 \propto e^{-(m_{T}-m_{\rho})/T} / [T(m_{\rho}+T)]$, where $T$ is set to be 317 MeV for having a $\langle p_{T} \rangle$ of 830 MeV~\cite{Adams:2003cc}, and $m_{\rho}$ is the rest mass of $\rho$-resonance. Note that the only available experimental data for $\rho$-resonance spectra at RHIC energies are measured for $40-80\%$ central Au+Au collisions at $\sqrt{s_{NN}} = 200 $ GeV~\cite{Adams:2003cc}, which does not match the $30-40\%$ centrality mentioned above. However, for a qualitative study this mismatch will not affect the conclusion. In this paper the finite event plane resolution is not taken into consideration in simulations. A finite event plane resolution will smear the difference between $x$ and $y$ directions, and make both $r_\mathrm{rest}$ and $r_\mathrm{lab}$, as well as the ratio of the two, approach unity. If needed this effect can be taken into account by smearing the reaction plane with well-established procedure~\cite{Poskanzer:1998yz}. In all simulations in this paper the reaction plane is assumed to be known exactly. The $dN/d\eta (dN/dy)$ distributions are taken to be flat in a range of [-1,1] for primordial pions ($\rho$ resonances). By default $\sim4$ million events are simulated for each data point in almost all figures of this section, except for Fig.~\ref{fig:a1Change_mostSimpleCase}, ~\ref{fig:primV2V3PtChange_typicalResFlow_compareA1} and ~\ref{fig:resV2V3PtChange_typicalPrimFlow_compareA1} in which $\sim10$ million events for each data point are simulated.

 \subsection{No backgrounds}\label{subsec:signalOnly}
 
\begin{figure}[htbp]
\centering
\makebox[1cm]{\includegraphics[width=0.45 \textwidth]{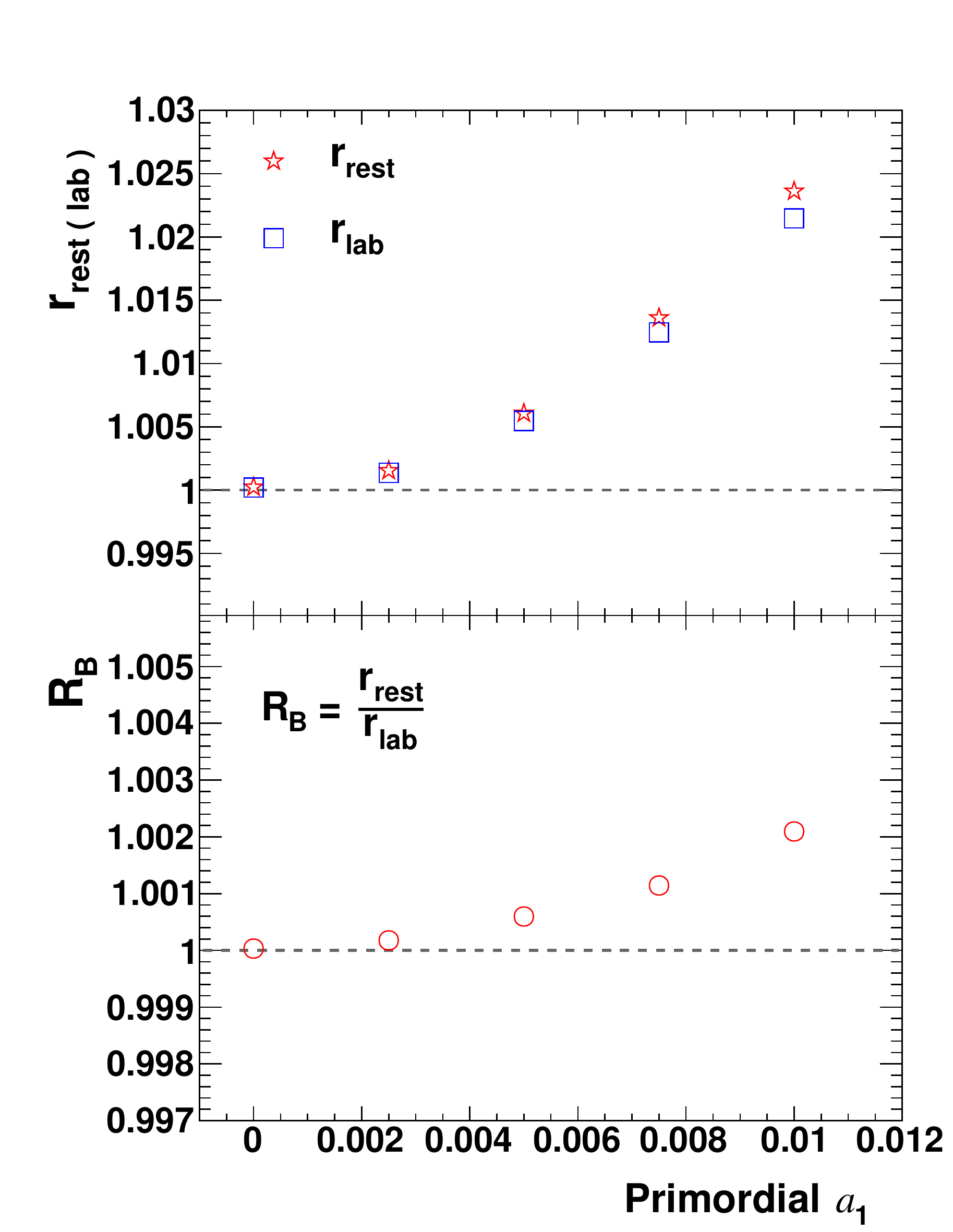}}
\caption{(Color online) $r_{\mathrm{rest}}$ and $r_{\mathrm{lab}}$(top panel), as well as $R_{B}$ (bottom panel) for a simple-case simulation in which only $a_{1}$ is introduced, with no backgrounds. }
\label{fig:a1Change_mostSimpleCase}
\end{figure}

Fig.~\ref{fig:a1Change_mostSimpleCase} presents a simulation with the CME signal only, and no backgrounds ($v_2$ and $v_3$ vanish for all particles, and no global spin alignment for resonances). In this simple case, $r_{\mathrm{rest}}$ and $r_{\mathrm{lab}}$ (top panel), as well as $R_{B}$ (bottom panel) are consistent with unity when $a_{1}=0$, and increase with increasing $a_{1}$. The deviation from unity for $R_{B}$ is about an order of magnitude smaller than that for $r_{\mathrm{rest}}$ and $r_{\mathrm{lab}}$, which is not a surprise as the additional sensitivity of $r_{\mathrm{rest}}$ over $r_{\mathrm{lab}}$ is a second-order effect. Indeed as $r_{\mathrm{lab}}$ and $r_{\mathrm{rest}}$ are visually very close to each other, for clarity reason in the rest of the paper it is chosen to show  only $r_{\mathrm{rest}}$ in figures.

\subsection{Resonance $v_{2}$ as fixed values}\label{subsec:ResonanceV2Change}

\begin{figure}[htbp]
\centering
\makebox[1cm]{\includegraphics[width=0.45 \textwidth]{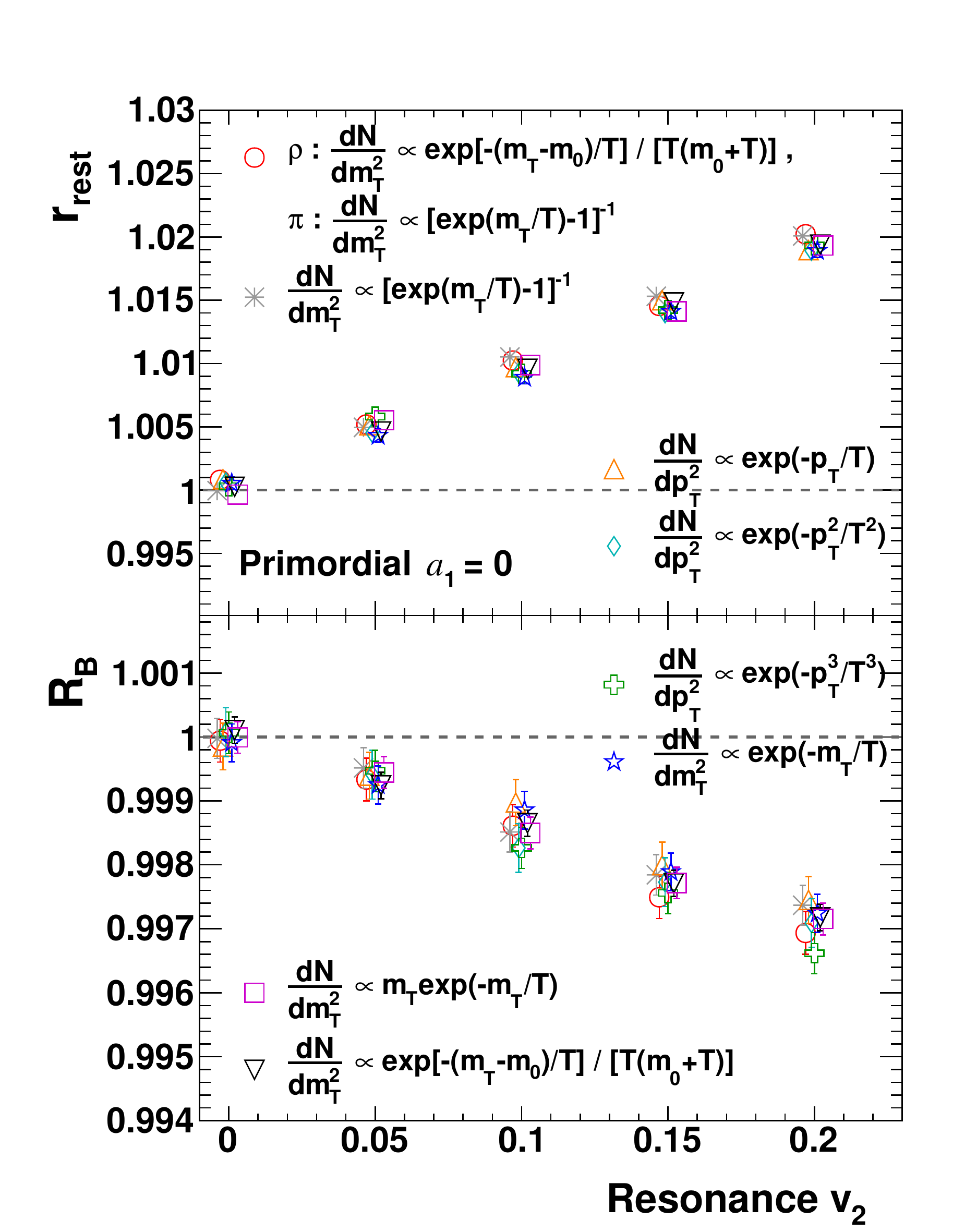}}
\caption{(Color online) $r_{\mathrm{rest}}$ and $R_{B}$ as a function of resonance $v_{2}$, for various transverse spectra. Formulae for spectra are from ~\cite{Abelev:2008ab} and ~\cite{Wang:2016iov}, with temperatures for each  are individually tuned to yield $\langle p_{T} \rangle$ of 400 MeV and 830 MeV for pions and $\rho$ resonances, respectively. Data points are shifted slightly in horizontal direction for clear view ( similar shift, when needed, has been applied for other plots in this paper ).}
\label{fig:resFixedV2Change_noPrimFlowNoA1_compareSpectra}
\end{figure}

In the simulation presented in Fig.~\ref{fig:resFixedV2Change_noPrimFlowNoA1_compareSpectra}, the elliptic flow for $\rho$ resonances is introduced. Here all resonances are generated according to same $v_{2}$ regardless of their $p_{T}$. Cases with $p_{T}$ dependent $v_{2}$ will be considered later in the paper. One can see that when there is no signal ($a_{1}=0$) and $v_{2}$ of resonance is the only background, $r_{\mathrm{rest}}$ increases with increasing resonance $v_{2}$ (top panel). It is known\cite{Feng:2018chm} that when resonances have in-plane elliptic flow, those with low $p_T$ (majority) tend to decay into large opening-angle pairs and result in more back-to-back pairs out-of-plane, mimicking a CME signal. In the rest frame although pairs that are originated from real resonance decays do not contribute to the apparent charge separation (barring finite global spin alignment which will be discussed later), pairs with one of them from resonance and another one picked randomly do. This will cause both $r_{\mathrm{rest}}$ and $r_{\mathrm{lab}}$ (not shown) to increase with increasing resonance $v_2$. 

The bottom panel of Fig.~\ref{fig:resFixedV2Change_noPrimFlowNoA1_compareSpectra} shows that $R_{B}$, on the contrary, decreases with increasing resonance $v_{2}$ -- an opposite trend than $r_{\mathrm{rest}}$. This pattern has been observed for all spectra formulae that can practically describe data ~\cite{Abelev:2008ab}. This can be explained as following: Inclusive particle pairs consist of those from resonance decays and those from other combinatorials. If there is no global spin alignment, the azimuthal distribution is isotropic for pairs from resonance decays but not necessarily so for those from other sources. That means by definition the azimuthal distribution of particles in pair's rest frame contains a fraction of isotropic source. Such fraction clearly does not contribute to the apparent charge separation, and one cannot obviously identify a similar fraction in the laboratory frame. One has to be reminded that the more symmetrical/isotropic the system is, the closer to unity the ratio $r_{\mathrm{rest(lab)}}$ is. Because of that, $r_{\mathrm{rest}}$ tends to be always closer to unity than $r_{\mathrm{lab}}$. That results in a relatively smaller slope of $r_{\mathrm{rest}}$ versus resonance $v_2$, and as a consequence $R_{B}$ decreases with increasing resonance $v_{2}$.

\begin{figure}[htbp]
\centering
\makebox[1cm]{\includegraphics[width=0.45 \textwidth]{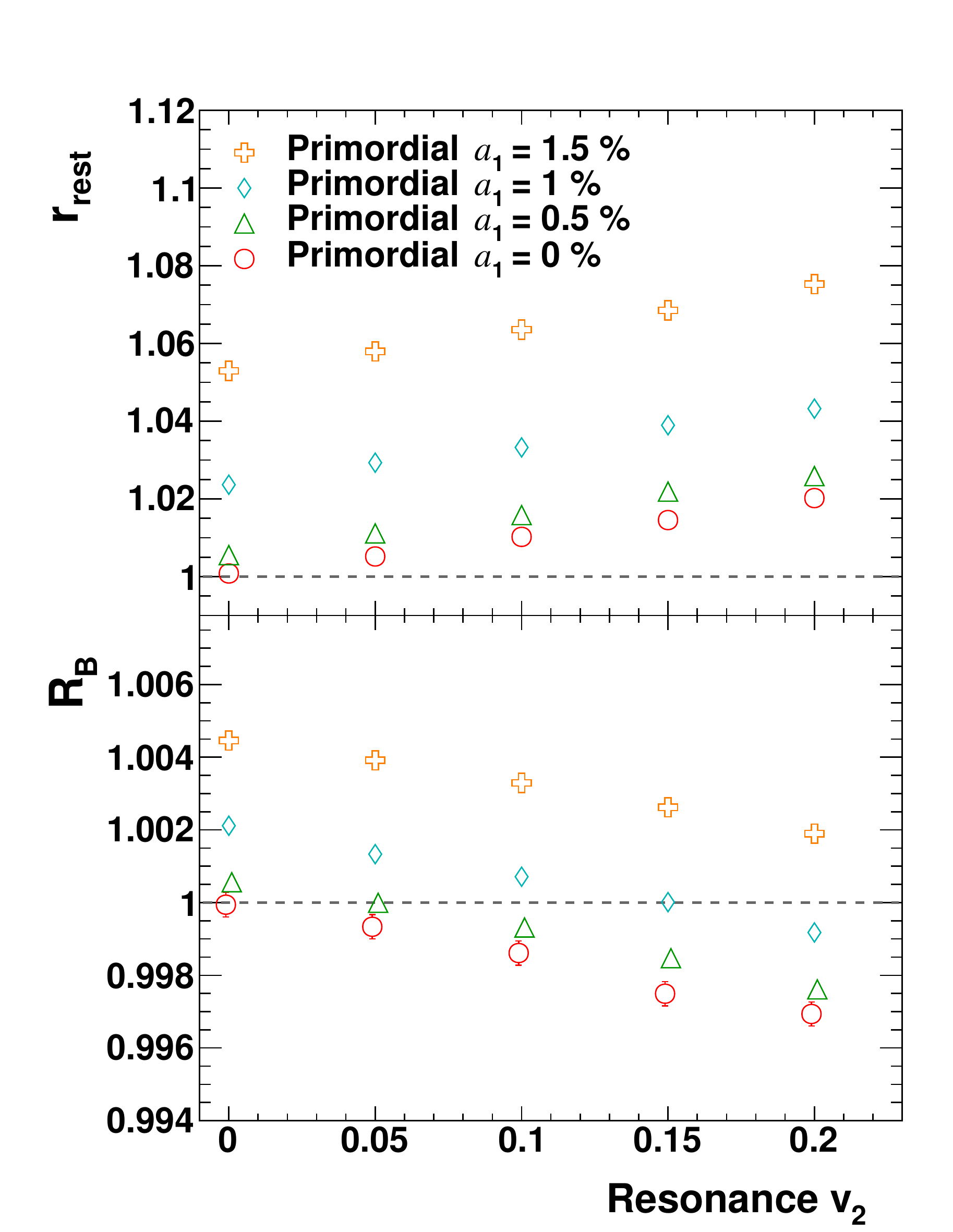}}
\caption{(Color online) $r_{\mathrm{rest}}$ and $R_{B}$ as a function of resonance $v_{2}$, for various $a_{1}$ values. }
\label{fig:resFixedV2Change_noPrimFlow_compareA1}
\end{figure}

In Fig.~\ref{fig:resFixedV2Change_noPrimFlow_compareA1}, a similar study is repeated with various $a_{1}$ introduced to primordial pions. The spectra of primordial pions and resonances are set to be the default setups as aforementioned at the beginning of section \ref{sec:ToyModelSimulation}. As expected, both $r_{\mathrm{rest}}$ and $R_{B}$ increase with increasing $a_{1}$, on top of values induced by resonance $v_{2}$ alone.

\subsection{Resonance $\rho_{00}$}\label{subsec:ResonanceRho00}

\begin{figure}[htbp]
\centering
\makebox[1cm]{\includegraphics[width=0.45 \textwidth]{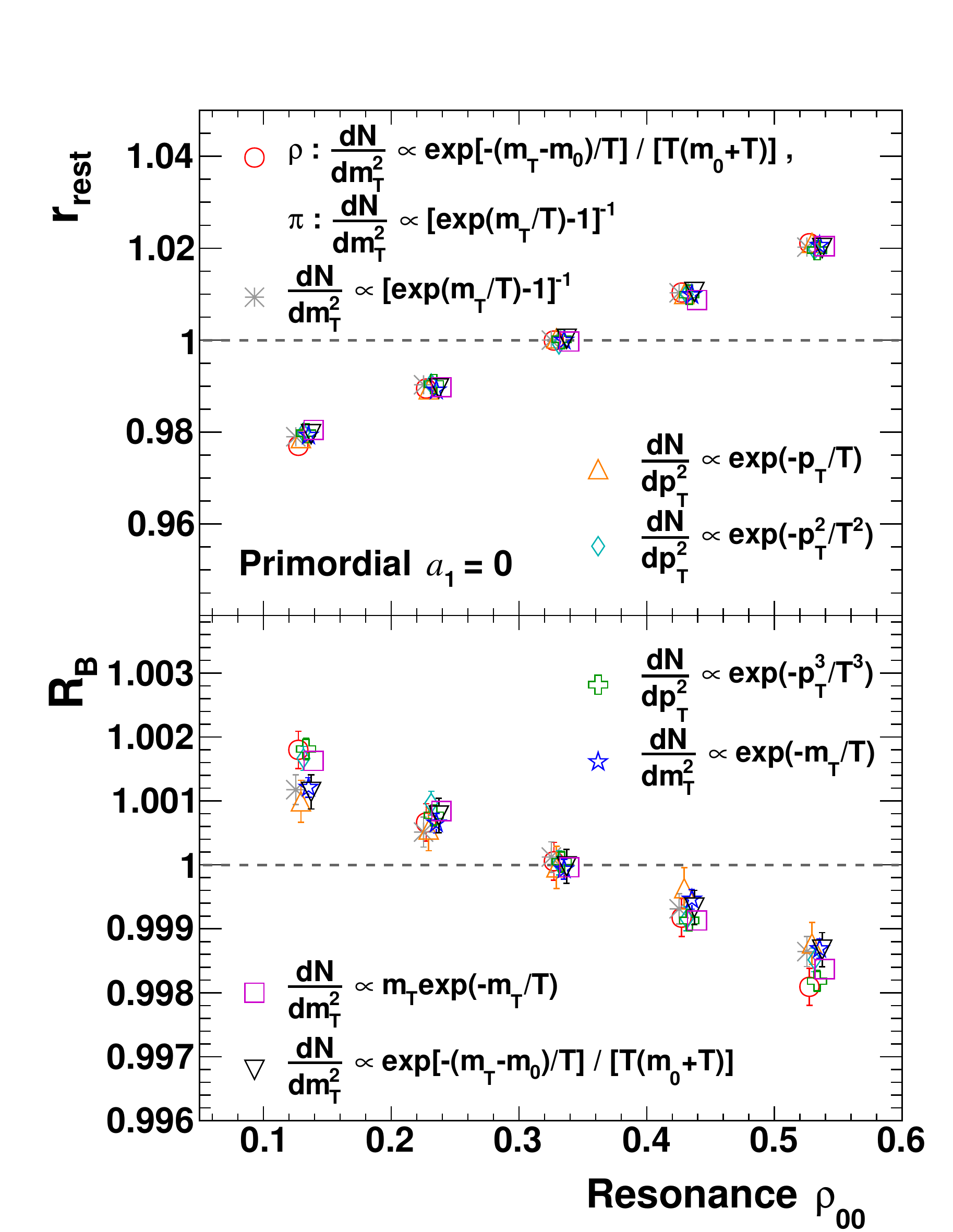}}
\caption{(Color online) $r_{\mathrm{rest}}$ and $R_{B}$ as a function of resonance $\rho_{00}$ for various transverse spectra. Choices of spectra are the same as in Fig.~\ref{fig:resFixedV2Change_noPrimFlowNoA1_compareSpectra}. }
\label{fig:rho00Change_noFlowNoA1_compareSpectra}
\end{figure}

\begin{figure}[htbp]
\centering
\makebox[1cm]{\includegraphics[width=0.45 \textwidth]{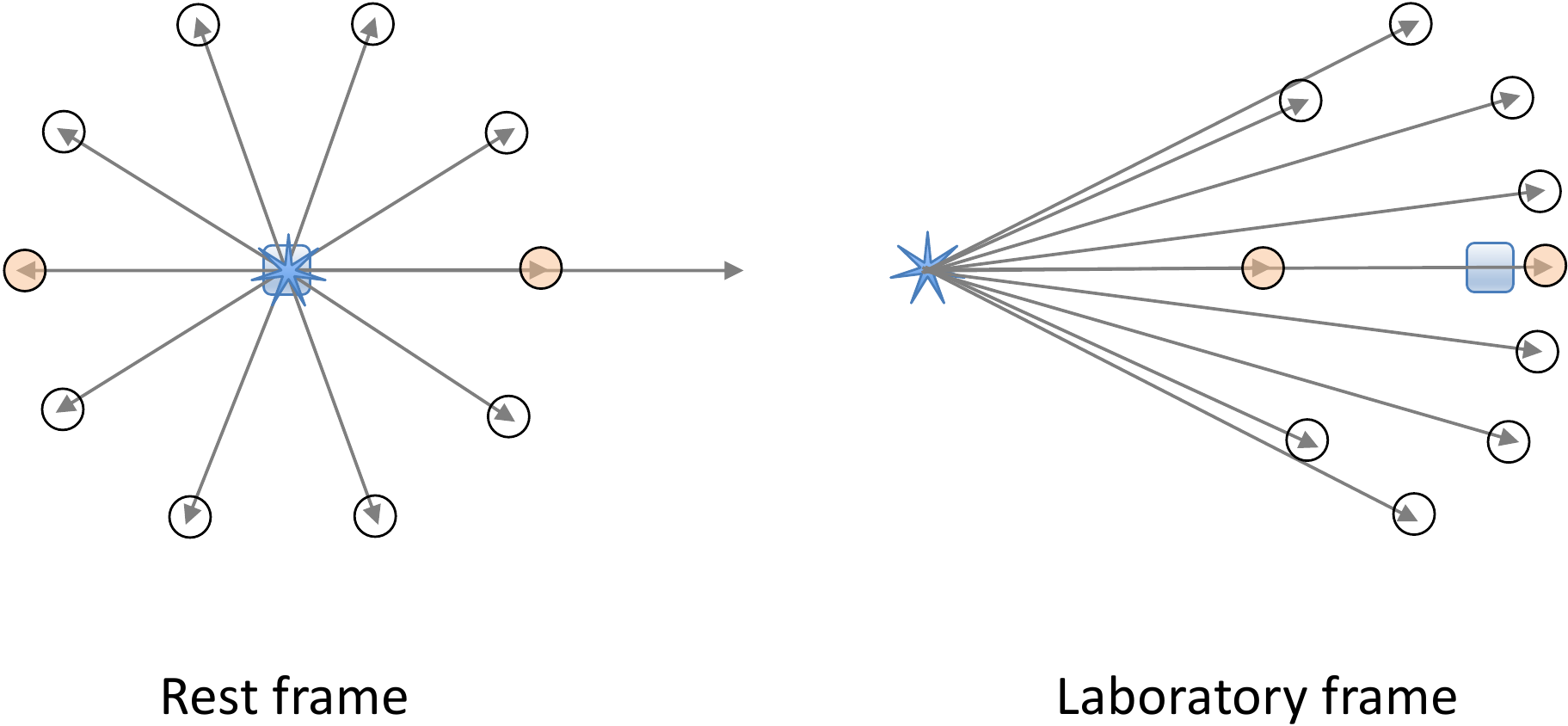}}
\caption{(Color online) When boost is strong a back-to-back pair (in orange color) in rest frame can become two particles traveling in the same direction in the laboratory frame, causing a depletion of back-to-back pairs along the boost direction.}
\label{fig:backToBackPairBoost-crop}
\end{figure}

It has been pointed that resonances with even spin can possess global spin alignment which tends to, in their rest frames, align two daughters either in the $y$ direction ($\rho_{00} > 1/3$), or in the $x-z$ plane ($\rho_{00} < 1/3$)~\cite{Liang:2004ph,Liang:2004xn,Liang:2007ma,Betz:2007kg,Gao:2007bc,Becattini:2013vja}. Considering the projection of many pairs onto the transverse plane only, loosely speaking the global spin alignment acts like ``elliptic flow" in the rest frame. For a reason similar to elliptic flow, the global spin alignment is also expected to cause an apparent charge separation. A $\rho_{00}$ larger (smaller) than 1/3  would mean more apparent charge separation in the $y$ ($x$) direction, causing both $r_{\mathrm{lab}}$ and $r_{\mathrm{rest}}$ to be larger (smaller) than unity. Such effect has not been discussed previously. In Fig.~\ref{fig:rho00Change_noFlowNoA1_compareSpectra}  $r_{\mathrm{rest}}$ and $R_{B}$ are shown as a function of resonance $\rho_{00}$ for various transverse spectra, with $a_{1} = 0$ and with no flow effects introduced anywhere. For the case of no global spin alignment ($\rho_{00} = 1/3$), $r_{\mathrm{rest}}$ and $R_{B}$ are at unity as they should. When there is global spin alignment ($\rho_{00} \neq 1/3 $), both ratios are not at unity anymore. $R_{B}$ is found to change in the opposite direction to $r_{\mathrm{rest}}$-change when responding to the change of $\rho_{00}$. This pattern holds again for all transverse spectra shapes that have been considered. This can be understood as the effect of Lorentz boost. When the boost is strong and the velocity of the pair's center of mass overcomes daughter's velocity in the rest frame,
for a fraction of resonances their daughter pairs, which are back-to-back in the rest frame, can become pairs going in the same direction in the laboratory frame, causing in the laboratory frame a depletion of back-to-back pairs in the direction of boost (see Fig.~\ref{fig:backToBackPairBoost-crop} for an illustration). 
 For such effect to happen, a resonance itself must move fast enough to overcome daughter's momentum along the boost direction. If, say, daughter's momentum projection in $x$ direction is smaller than that in $y$ direction, then $x$ direction is more vulnerable to depletion, causing an extra, apparent charge separation along $y$ direction in the laboratory frame. This is indeed the case for finite global spin alignment: as  $\rho_{00} \neq 1/3$ means in the rest frame daughter's momentum projection in the $x$ and $y$ direction are unbalanced, thus they have unequal vulnerability to depletion.

\begin{figure}[htbp]
\centering
\makebox[1cm]{\includegraphics[width=0.45 \textwidth]{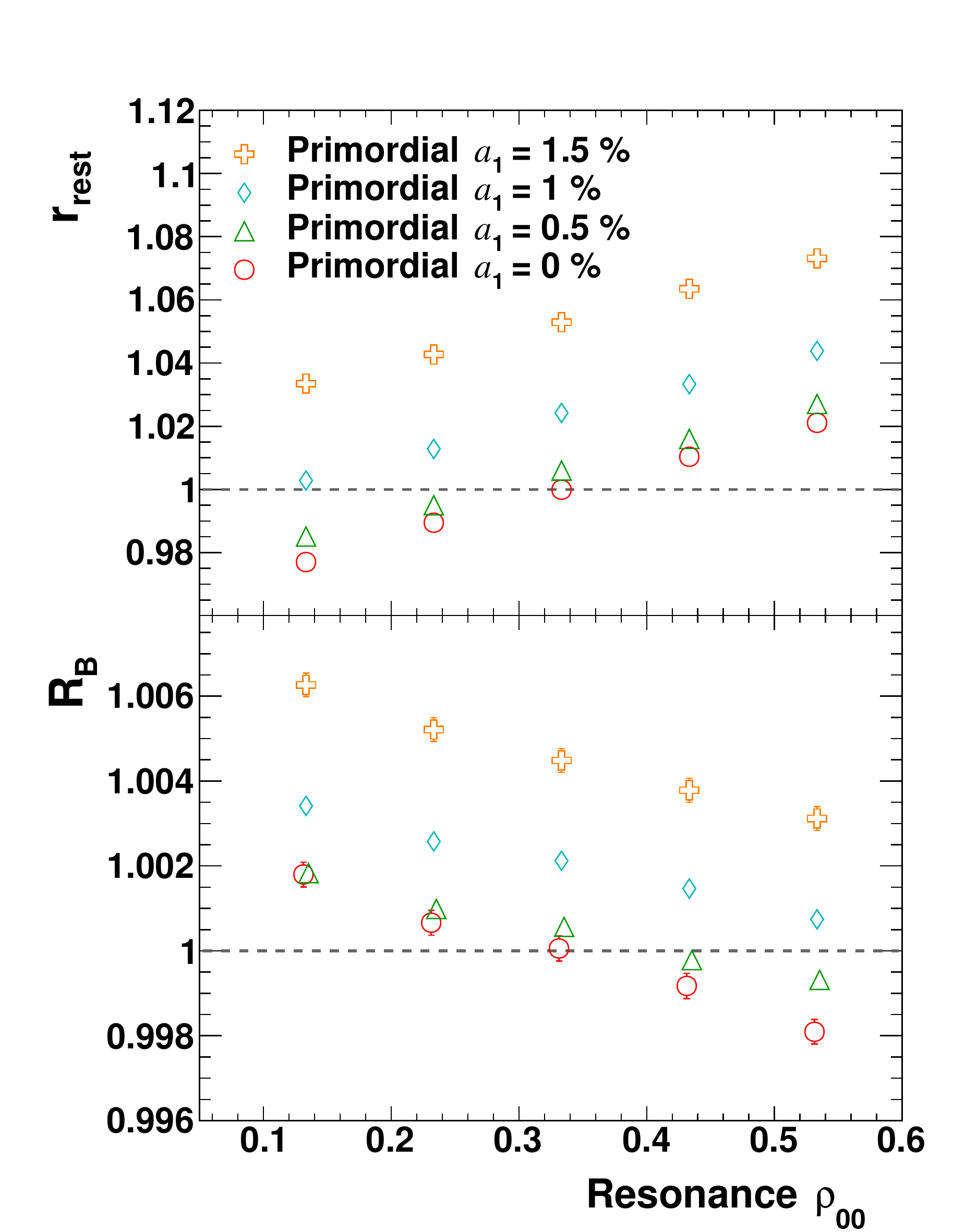}}
\caption{(Color online) $r_{\mathrm{rest}}$ and $R_{B}$ as a function of resonance $\rho_{00}$, for various $a_{1}$ values. No flow effects are included.}
\label{fig:rho00Change_noFlow_compareA1}
\end{figure}

In Fig.~\ref{fig:rho00Change_noFlow_compareA1}, similar studies are repeated with various $a_{1}$ introduced to primordial pions, with spectra of primordial pions and resonances set to be default ones as mentioned in the beginning of section \ref{sec:ToyModelSimulation}. As expected, both $r_{\mathrm{rest}}$ and $R_{B}$ increase with increasing $a_{1}$, on top of values induced by resonance $\rho_{00}$ alone.

 Note in this study a wide $\rho_{00}$ range is chosen in order to clearly identify/demonstrate the pattern, which may have exaggerated the situation. Experimentally $\rho_{00}$ has been studied for $\phi$-meson for Au+Au collisions at $\sqrt{s_{NN}} = 200 $ GeV~\cite{Zhou:2019lun}, and it is found to be less than 0.38 for $p_{T} > 1.2$  GeV/$c$. Preliminary $\rho_{00}$ measurements for $K^{*0}$-meson are found to be smaller than 1/3, at both RHIC and LHC energies~\cite{Singh:2019QM, Kundu:2019QM}. So far there is no experimental guidance on $\rho_{00}$ for $\rho$-resonance, and our study calls for such measurements. 

\subsection{$p_{T}$ dependent $v_{2}$ and $v_{3}$ of primordial pions}\label{subsec:ptDependentV2Prim}

In this subsection and the following subsections, how the two observables respond to realistic flow effects are studied. The NCQ-inspired function~\cite{Dong:2004ve} is used to introduce elliptic flow for primordial pions and $\rho$ resonances,
\begin{eqnarray}
\begin{aligned}
v_{2}/n = \mathbf{a}/(1+e^{-[(m_{T}-m_{0})/n-\mathbf{b}]/\mathbf{c}}) - \mathbf{d},
\end{aligned}
\label{eq:NCQ_v2}
\end{eqnarray}
where $n=2$ is the number of constituent quarks. By default parameters $\mathbf{a}$, $\mathbf{b}$, $\mathbf{c}$ and $\mathbf{d}$ take same values as in ~\cite{Wang:2016iov} for $30-40\%$ central Au+Au collisions at $\sqrt{s_{NN}} = 200 $ GeV. Unless otherwise specified, $v_{3}$ at any given $p_{T}$ is set to be $1/5$ of corresponding $v_{2}$~\cite{Shou:2014cja}, for both primordial poins and $\rho$ resonances. No spin alignment is introduced for $\rho$ resonances. The study with realistic flow together with global spin alignment will be presented in a later subsection.

\begin{figure}[htbp]
\centering
\makebox[1cm]{\includegraphics[width=0.45 \textwidth]{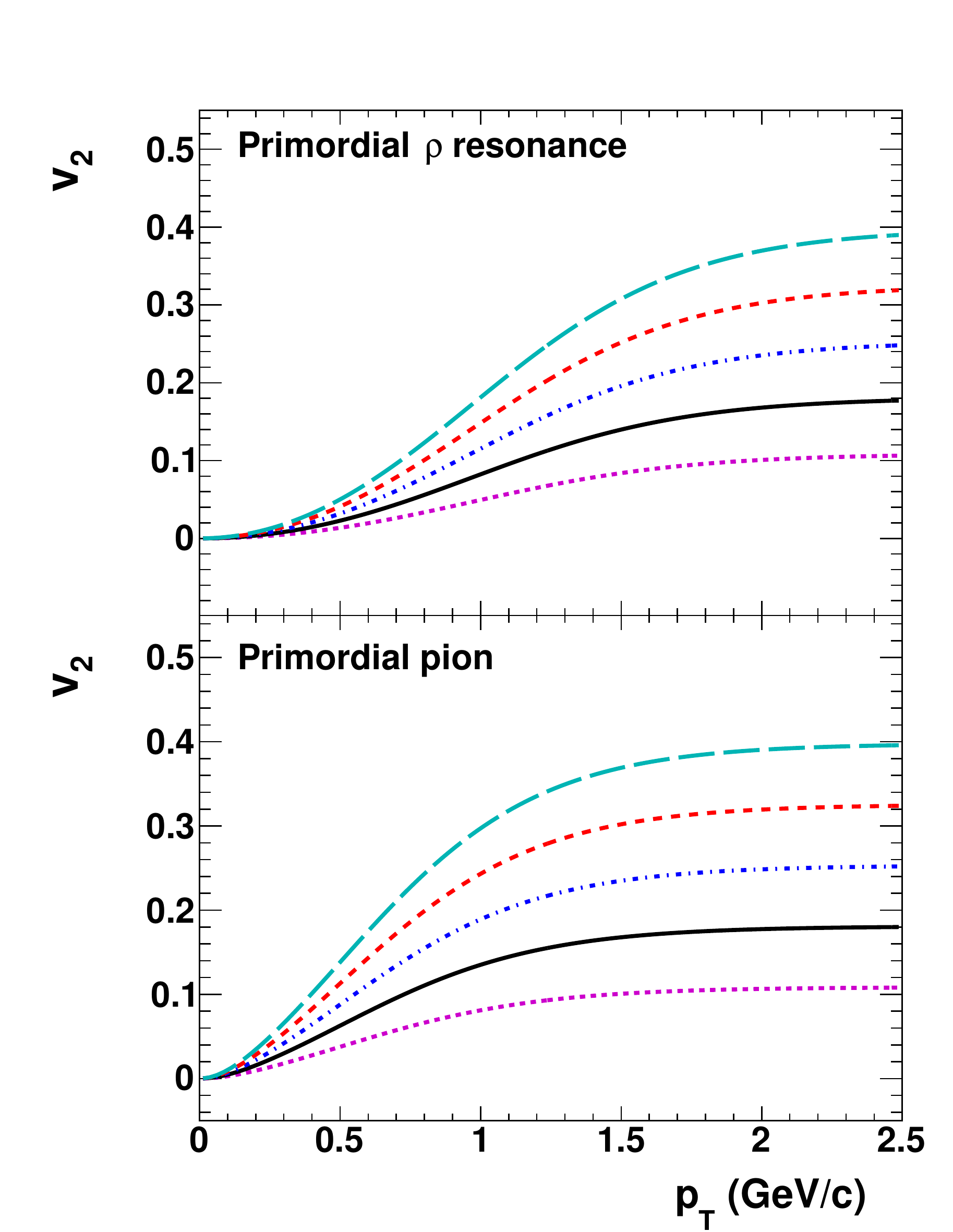}}
\caption{(Color online) $v_{2}(p_{T})$ implemented in simulations, for $\rho$ resonances (top panel) and primordial pions (bottom panel). Within each panel, $\mathbf{a}$-parameter values in Eq. (\ref{eq:NCQ_v2}) are, from top to bottom, 0.275, 0.225, 0.175, 0.125 and 0.075, respectively. For each panel the curve with solid black line corresponds to the case with default value ($0.125$) of $\mathbf{a}$-parameter  taken from ~\cite{Wang:2016iov}. }
\label{fig:plotV2PtAParChangeDemo}
\end{figure}
\begin{figure}[htbp]
\centering
\makebox[1cm]{\includegraphics[width=0.45 \textwidth]{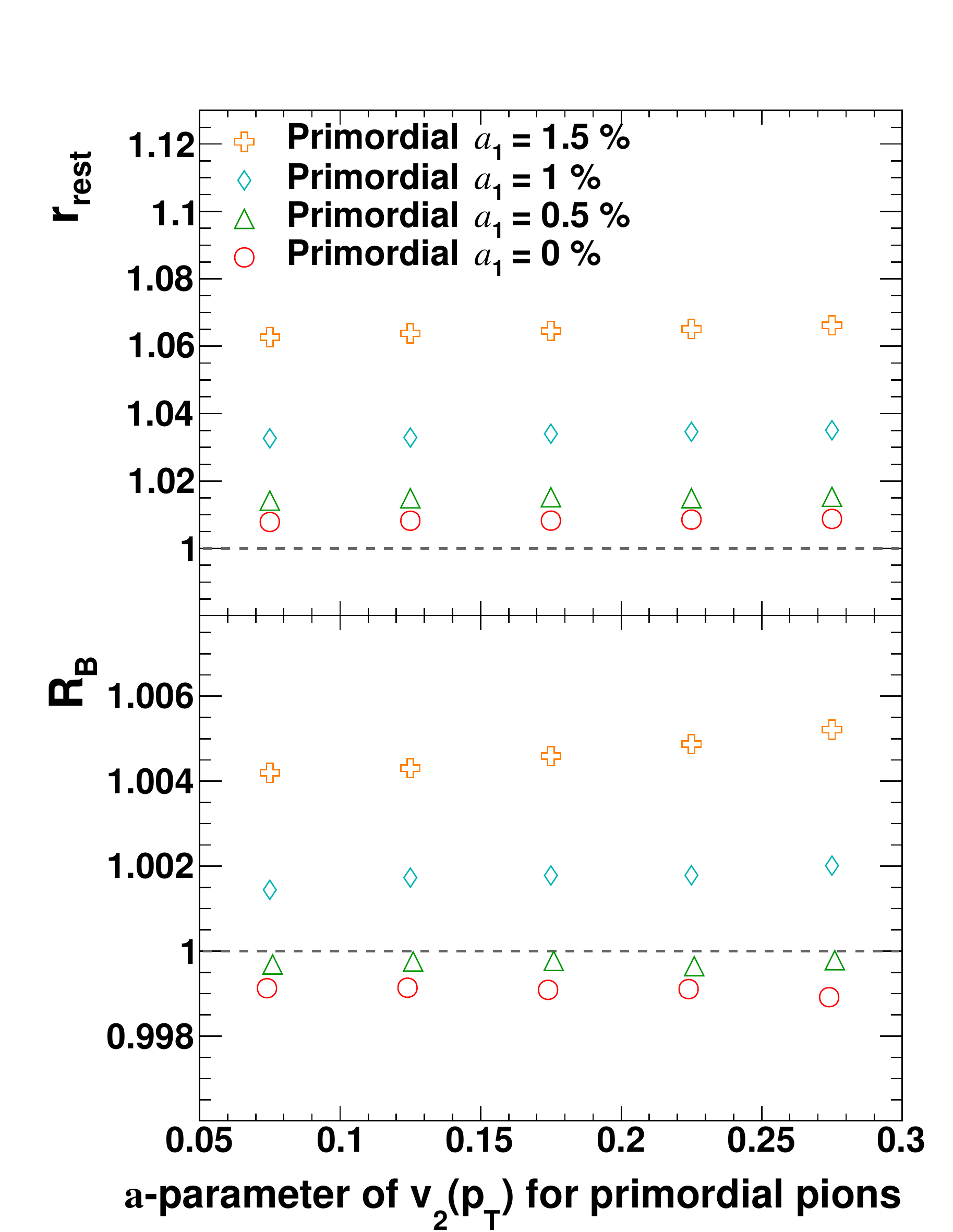}}
\caption{(Color online) $r_{\mathrm{rest}}$ and $R_{B}$ as a function of $\mathbf{a}$-parameter of primordial pions in $v_{2}(p_{T})$ description, for various $a_{1}$ values. }
\label{fig:primV2V3PtChange_typicalResFlow_compareA1}
\end{figure}

The parameter $\mathbf{a}$ in Eq. (\ref{eq:NCQ_v2}) is varied to change $v_{2}(p_{T})$, and the parameter $\mathbf{d}$ has been adjusted accordingly to enssure $v_{2} = 0$ when $p_{T} = 0$. The effect of this variation on $v_{2}$ is illustrated in Fig.~\ref{fig:plotV2PtAParChangeDemo}. 
In Fig.~\ref{fig:primV2V3PtChange_typicalResFlow_compareA1} $r_{\mathrm{rest}}$ and $R_{B}$ are presented as a function of $\mathbf{a}$-parameter of primordial pions. Note that in this study $v_{3}$ for primordial pions also changes with $\mathbf{a}$-parameter, as $v_{3}$ at any given $p_{T}$ has been set to be 1/5 of corresponding $v_{2}$.  $v_{2}(p_{T})$ and $v_{3}(p_{T})$ for $\rho$ resonances are introduced according to their aforementioned default configurations and are kept unchanged. 

One would find that when there is no CME-induced charge separation ($a_{1}=0$), $r_{\mathrm{rest}}$ and $R_{B}$ are at opposite sides of unity, and this is largely due to the presence of finite $v_{2}$ of $\rho$-resonance. When there is finite $a_{1}$, $r_{\mathrm{rest}}$ increases slightly with the $\mathbf{a}$-parameter, and surprisingly $R_{B}$ also increases slightly with increasing $\mathbf{a}$-parameter. This has to be caused by a combination of finite $\rho$-resonance $v_{2}$ and the change of $v_{2}$ of primordial pions. However, the reasoning for it at microscopic, dynamical level is not obvious for the moment and is a subject of future study. 

\begin{figure}[htbp]
\centering
\makebox[1cm]{\includegraphics[width=0.45 \textwidth]{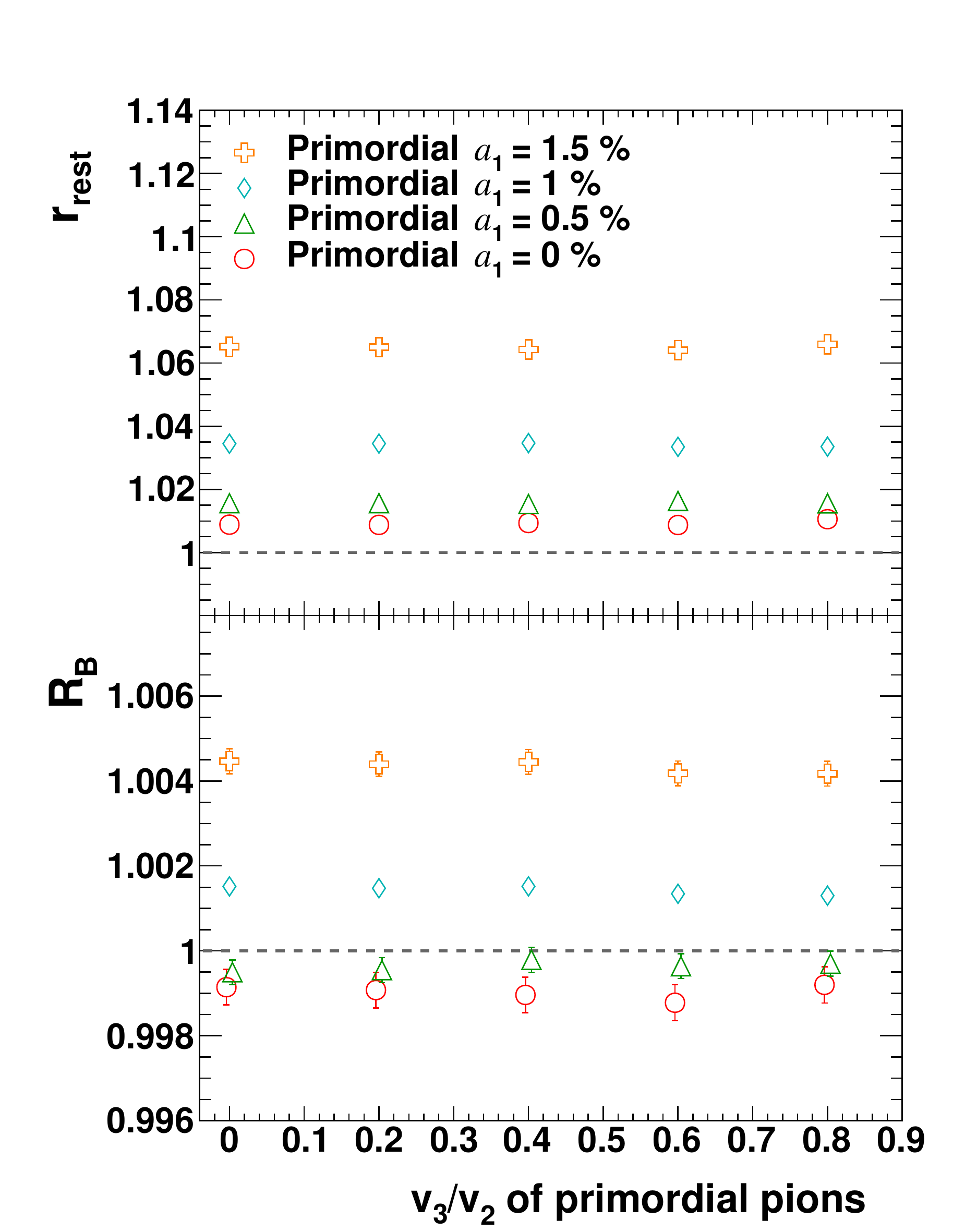}}
\caption{(Color online) $r_{\mathrm{rest}}$ and $R_{B}$ as a function of $v_{3}(p_{T})/v_{2}(p_{T})$ ratio of primordial pions, for various $a_{1}$ values. }
\label{fig:primV3Change_typicalPrimV2TypicalResFlow_compareA1}
\end{figure}

In Fig.~\ref{fig:primV3Change_typicalPrimV2TypicalResFlow_compareA1}, the effect of $v_{3}$ alone is studied by varying $v_{3}(p_{T})$ of primordial pions while keeping everything else unchanged. This is implemented by setting $v_{3}$ to be a fraction, which itself varies, of $v_{2}$ everywhere in $p_{T}$, while keeping $v_{2}(p_{T})$ unchanged. Both $r_{\mathrm{rest}}$ and $R_{B}$ are presented as function of ratio of $v_{3}/v_{2}$. As a reminder the case that is close to data is with $v_{3}/v_{2} = 0.2$. No obvious dependence on the change of $v_{3}$ of primordial pions can be seen.

\subsection{$p_{T}$ dependent $v_{2}$ and $v_{3}$ of $\rho$ resonances}\label{subsec:ptDependentV2Res}

\begin{figure}[htbp]
\centering
\makebox[1cm]{\includegraphics[width=0.45 \textwidth]{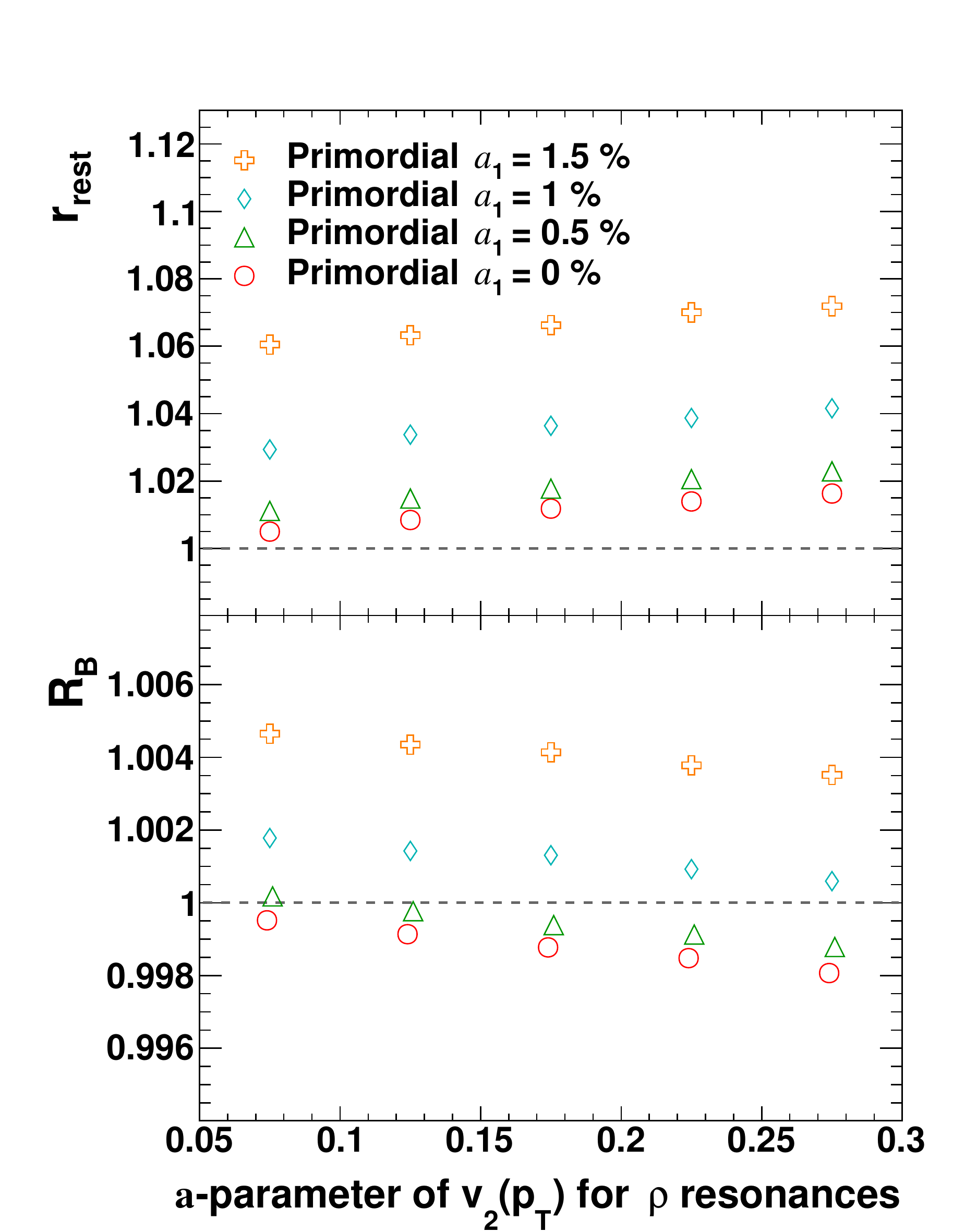}}
\caption{(Color online) $r_{\mathrm{rest}}$ and $R_{B}$ as a function of $\mathbf{a}$-parameter of $\rho$ resonances in $v_{2}(p_{T})$ description, for various $a_{1}$ values. }
\label{fig:resV2V3PtChange_typicalPrimFlow_compareA1}
\end{figure}

In this subsection studies similar to those in the previous subsection are repeated, but instead of varying the flow of primordial pions, here the flow of $\rho$ resonances are varied, while the flow of primordial pions are kept unchanged with their default configuration. No spin alignment is introduced for $\rho$-resonance. 

 Fig.~\ref{fig:resV2V3PtChange_typicalPrimFlow_compareA1} shows $r_{\mathrm{rest}}$ and $R_{B}$ as a function of $\mathbf{a}$-parameter of $\rho$ resonances. When there is no CME-induced separation, $r_{\mathrm{rest}}$ and $R_{B}$ deviate from unity in opposite directions. With a finite $a_{1}$, both observables increase on top of the values for the case of $a_{1} = 0$, and the pattern that $r_{\mathrm{rest}}$ and $R_{B}$ respond in opposite directions to the change of $\mathbf{a}$-parameter can be seen for all $a_{1}$ values.
 
\begin{figure}[htbp]
\centering
\makebox[1cm]{\includegraphics[width=0.45 \textwidth]{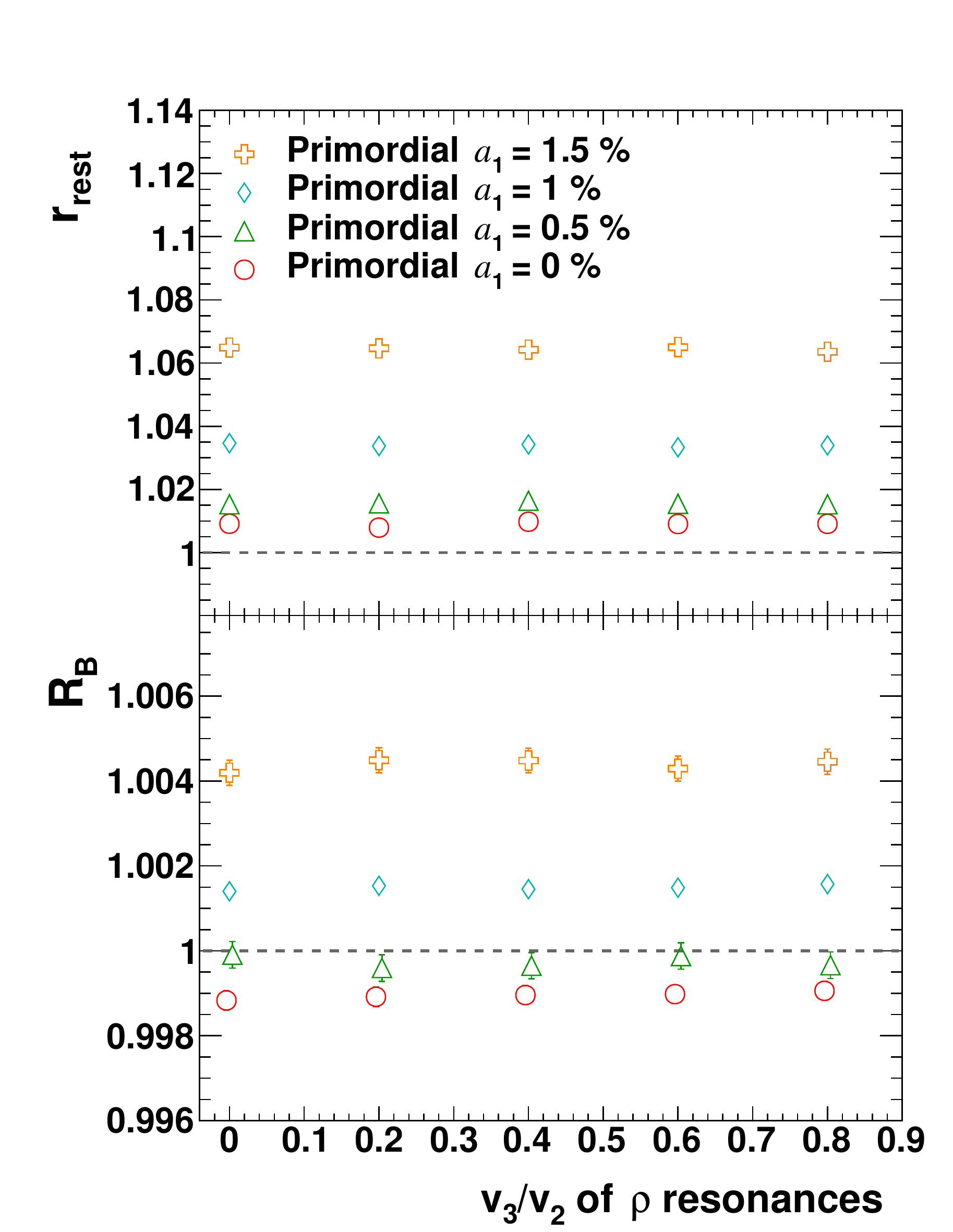}}
\caption{(Color online) $r_{\mathrm{rest}}$ and $R_{B}$ as a function of  $v_{3}(p_{T})/v_{2}(p_{T})$ ratio of $\rho$ resonances, for various $a_{1}$ values. }
\label{fig:resV3Change_typicalResV2TypicalPrimFlow_compareA1}
\end{figure}

In Fig.~\ref{fig:resV3Change_typicalResV2TypicalPrimFlow_compareA1}, following a similar procedure in Fig.~\ref{fig:primV3Change_typicalPrimV2TypicalResFlow_compareA1}, the resonance $v_{3}(p_{T})$ is varied while everything else is kept unchanged. Like the case for $v_{3}$ of primordial pions (Fig.~\ref{fig:primV3Change_typicalPrimV2TypicalResFlow_compareA1}), there is no noticeable effect due to $v_{3}$ change of $\rho$ resonances.

\subsection{Resonance $\rho_{00}$ together with $p_{T}$ dependent $v_{2}$ \& $v_{3}$ of primordial pions and $\rho$ resonances }\label{subsec:rho00WithTypicalFlow}

\begin{figure}[htbp]
\centering
\makebox[1cm]{\includegraphics[width=0.45 \textwidth]{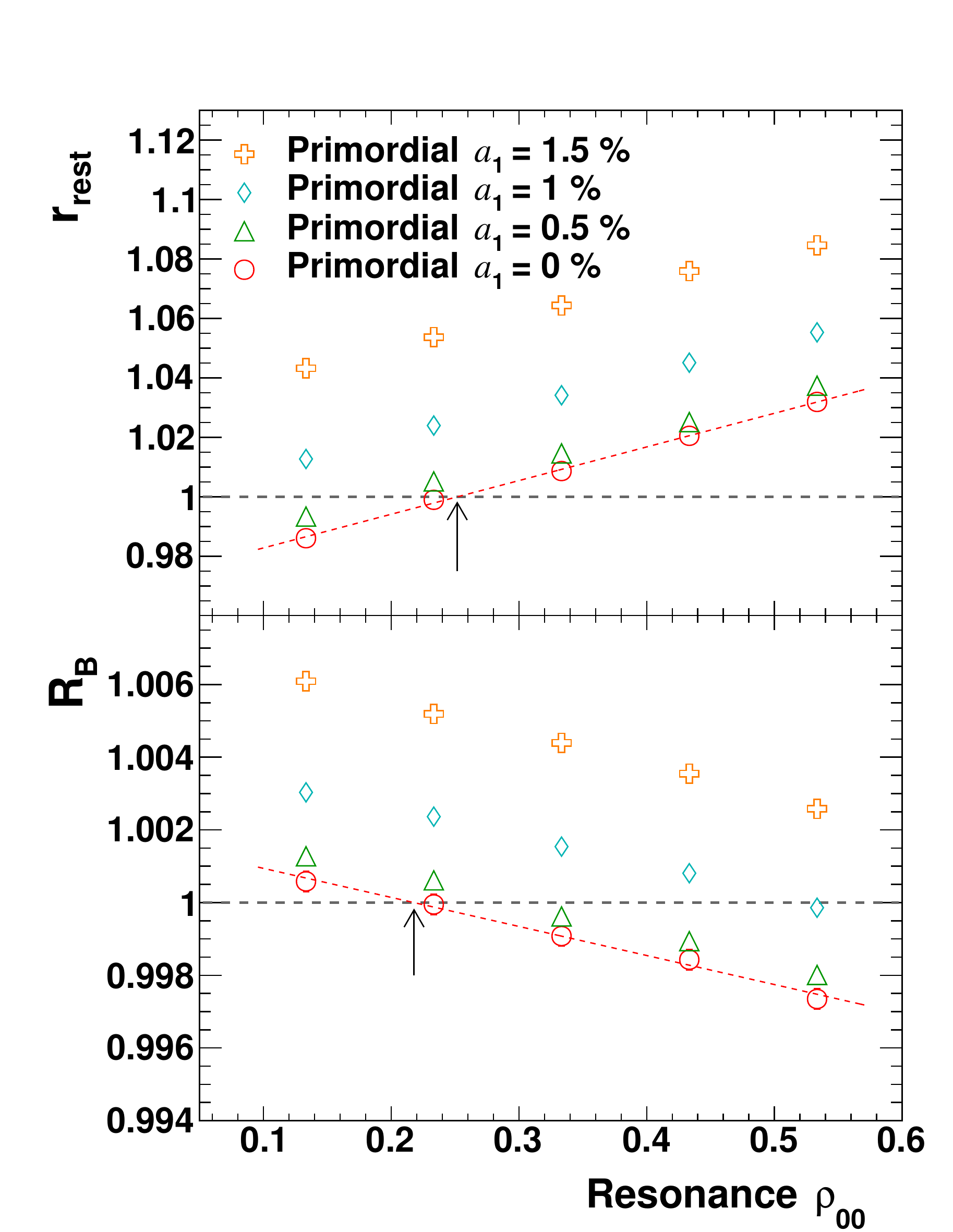}}
\caption{(Color online) $r_{\mathrm{rest}}$ and $R_{B}$ as a function of resonance $\rho_{00}$, for various $a_{1}$ values. Realistic flow effects have been included for both primordial pions and $\rho$ resonances. Arrows indicate the place where ratios cross unity for the case of $a_1 = 0$.}
\label{fig:rho00Change_typicalPrimAndResV2PtV3Pt_compareA1}
\end{figure}

In this subsection the $\rho_{00}$ study in Fig.~\ref{fig:rho00Change_noFlow_compareA1} is repeated, but instead of having no flow effects, here $p_{T}$ dependent flow effects, $v_{2}(p_{T})$ and $v_{3}(p_{T})$, are included for both primordial pions and $\rho$ resonances according to the aforementioned configuration in section~\ref{subsec:ptDependentV2Prim}. One can see (Fig.~\ref{fig:rho00Change_typicalPrimAndResV2PtV3Pt_compareA1}) again that $r_{\mathrm{rest}}$ and $R_{B}$ change in opposite directions when responding to $\rho_{00}$ change. Unlike in Fig.~\ref{fig:rho00Change_noFlow_compareA1}, here both observables are not at unity for $\rho_{00} = 1/3$ due to the presence of flow effects. 

Note that for the case without the CME (dashed line, with $a_1 = 0$), $r_{\mathrm{rest}}$ crosses unity at a larger $\rho_{00}$ value than $R_B$ (indicated by two arrows in Fig.~\ref{fig:rho00Change_typicalPrimAndResV2PtV3Pt_compareA1}). This crossing-ordering is due to the presence of in-plane resonance $v_2$. To understand this, imagine one starts with a system whose resonances have $\rho_{00} < 1/3$ but no $v_2$. According to section \ref{subsec:ResonanceRho00}, for this setup $r_{\mathrm{rest}} < 1$ and $R_B > 1$. From what is learned in section \ref{subsec:ResonanceV2Change}, an introduction of positive resonance $v_2$ to the system will increase $r_{\mathrm{rest}}$. Suppose the right amount of resonance $v_2$ is introduced so that $r_{\mathrm{rest}}$ reaches unity. Please be reminded that $r_{\mathrm{rest}} = 1$ means that pairs in the rest frame is symmetrical between $x$ and $y$ direction. However, for this system, if boosted back to laboratory frame, it will become a system with more apparent separation in $y$ direction due to the extra boost in $x$ direction to account for the in-plane flow. This, can be loosely imagined as that, a round shape in rest frame will become elongated in $y$ direction when boosted (with extra-boost in $x$ direction) back to laboratory frame. As a consequence, for a $\rho_{00}$ value at which $r_{\mathrm{rest}}$ crosses unity, $r_{\mathrm{lab}} > r_{\mathrm{rest}}$ and $R_B < 1$.  This feature ensures that, with backgrounds arising from resonance flow and global spin alignment, without the presence of the CME one cannot identify any $\rho_{00}$ at which $r_{\mathrm{rest}}$ and $R_B$ are above unity simultaneously.  

\begin{figure}[htbp]
\centering
\makebox[1cm]{\includegraphics[width=0.45 \textwidth]{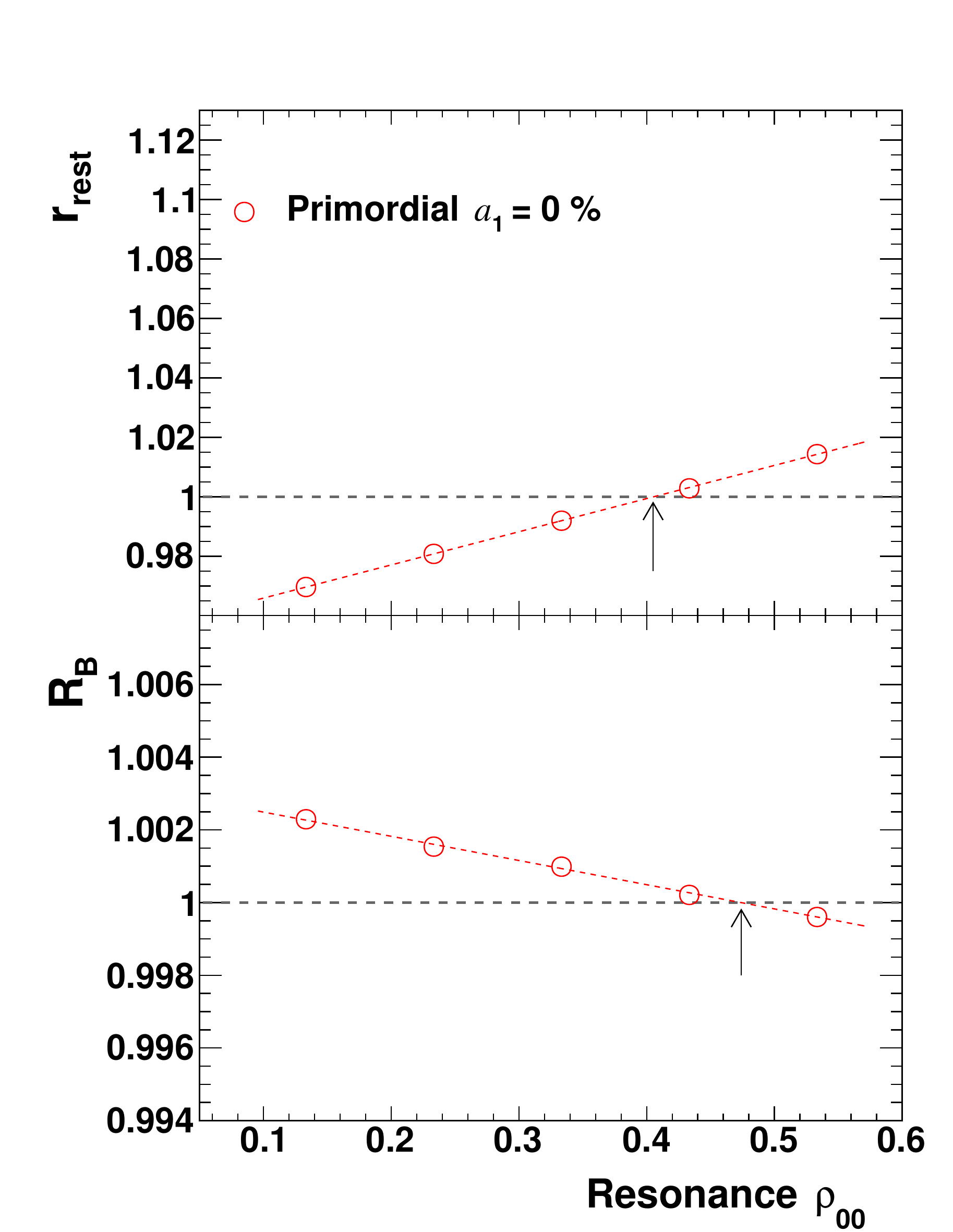}}
\caption{(Color online) $r_{\mathrm{rest}}$ and $R_{B}$ as a function of resonance $\rho_{00}$, for the $a_{1} = 0$ case in Fig.~\ref{fig:rho00Change_typicalPrimAndResV2PtV3Pt_compareA1} but with the sign of resonance $v_2$ set as negative. Arrows indicate the places where ratios cross unity.}
\label{fig:rho00Change_typicalPrimResV2PtV3PtNegResV2PtV3Pt_zeroA1}
\end{figure}

To consolidate this point, we repeat the simulation of the case of $a_1 = 0$ in Fig.~\ref{fig:rho00Change_typicalPrimAndResV2PtV3Pt_compareA1}, but with the sign of resonance $v_2$ flipped to negative (out-of-plane flow). This is a totally unphysical case but it is useful for testing the reasonings.  Following the arguement above, the crossing-ordering should be reversed. One can see that (Fig.~\ref{fig:rho00Change_typicalPrimResV2PtV3PtNegResV2PtV3Pt_zeroA1}) when the resonance flow is out-of-plane, the crossing-ordering is indeed reversed, consistent with the expectation from reasonings mentioned above.

\subsection{Resonance $p_{T}$}\label{subsec:resonancePt}

\begin{figure}[htbp]
\centering
\makebox[1cm]{\includegraphics[width=0.45 \textwidth]{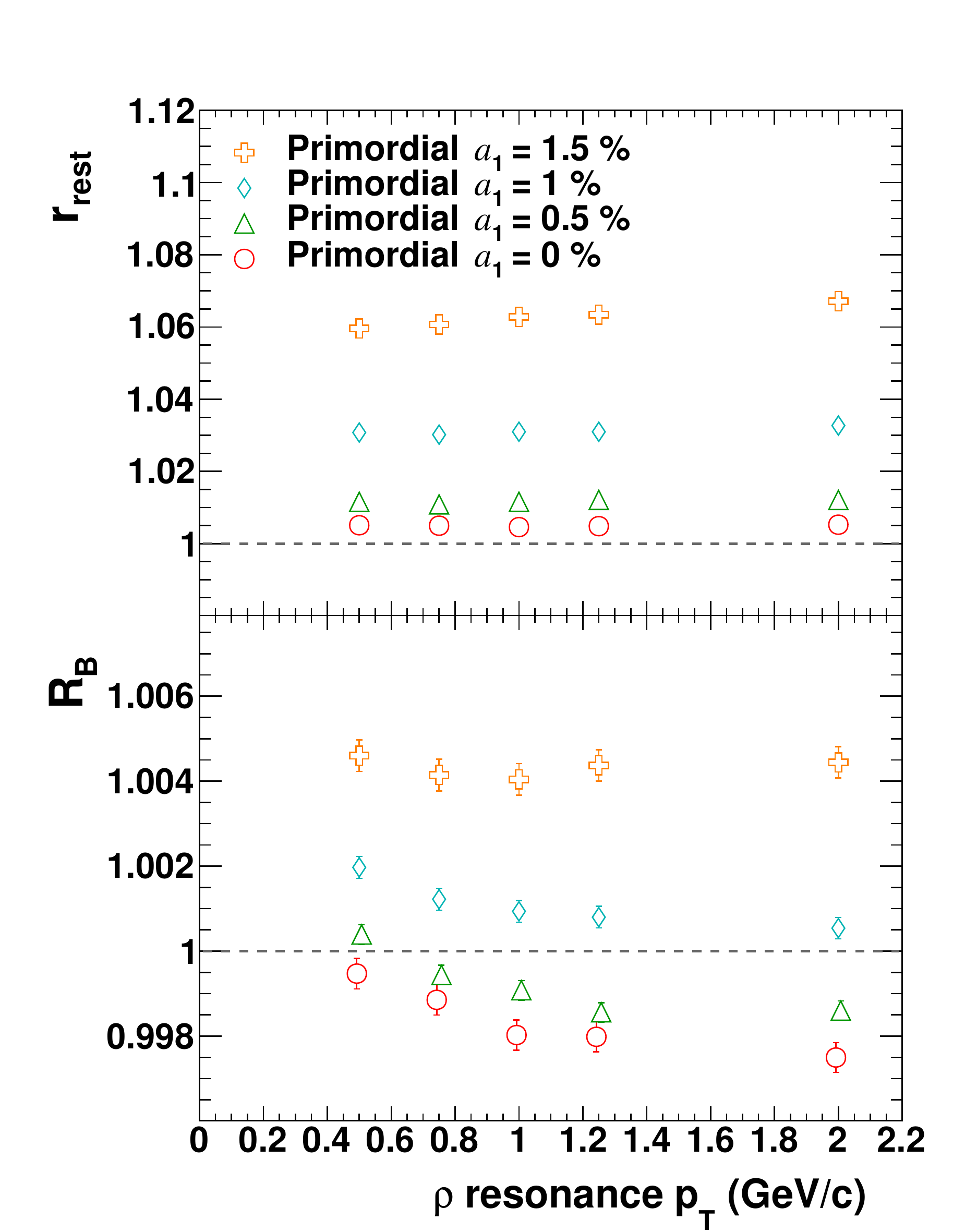}}
\caption{(Color online) $r_{\mathrm{rest}}$ and $R_{B}$ as a function of $\rho$ resonance $p_{T}$ as a fixed value.}
\label{fig:resFixedPtChange_typicalPrimFlowResFixed6PctV2NoResV3_compareA1}
\end{figure}

In a recent publication~\cite{Feng:2018chm}, it is pointed out that when acting together with resonance elliptic flow, low $p_{T}$ resonances have a tendency of emitting two daughters preferentially more perpendicular to the reaction plane than high $p_{T}$ resonances because of the large decay opening angle, while high $p_{T}$ resonances tend to emit two daughters close to each other and preferentially close to the reaction plane. Both effects will influence the fluctuation in $x$- and $y$-direction and should be considered as background in the CME-related analysis.  

\begin{figure}[htbp]
\centering
\makebox[1cm]{\includegraphics[width=0.45 \textwidth]{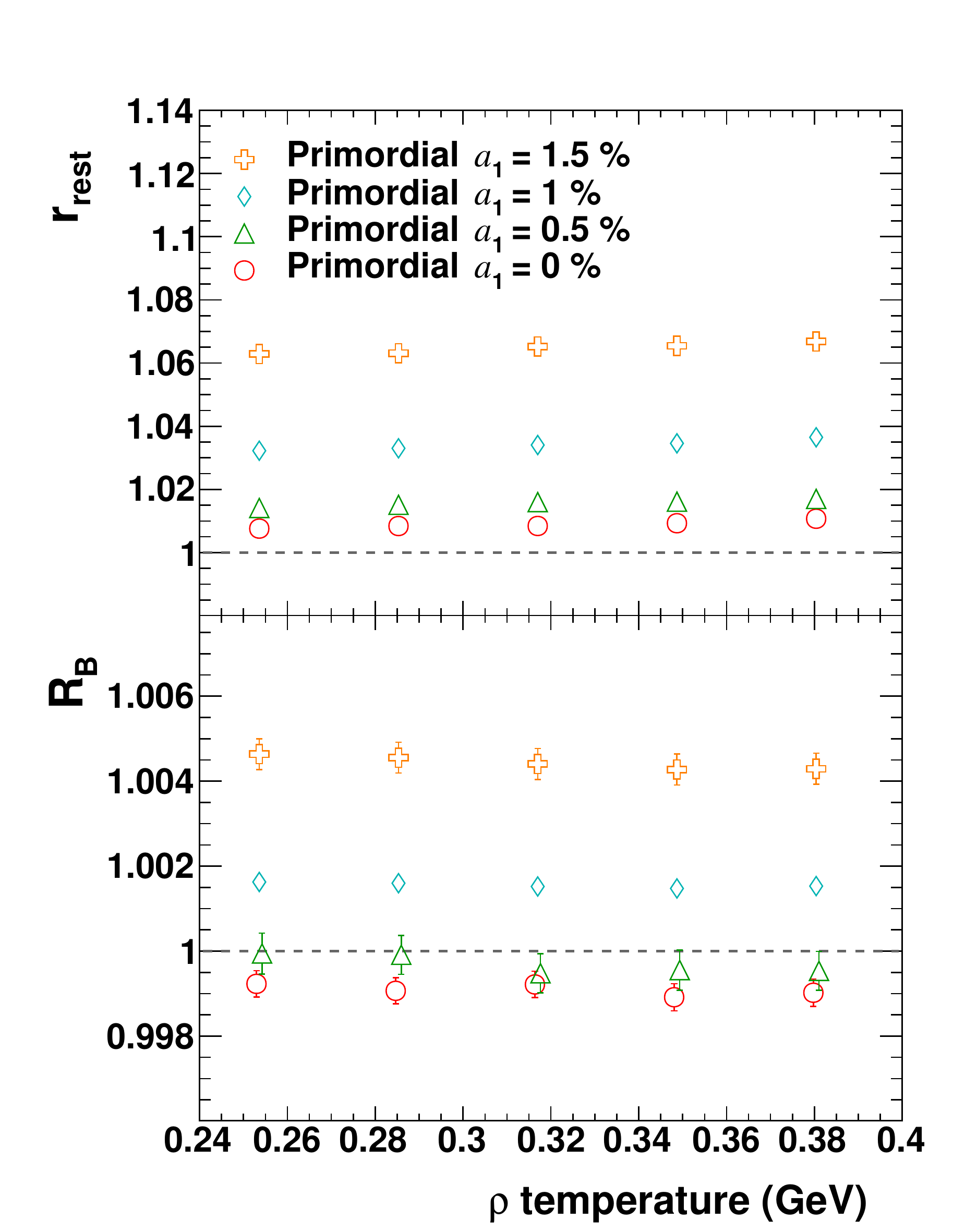}}
\caption{(Color online) $r_{\mathrm{rest}}$ and $R_{B}$ as a function of temperature of the transverse spectra of  $\rho$-resonance.}
\label{fig:resTChange_typicalPrimResFlow_compareA1}
\end{figure}

To repeat such a study for our observables, $\rho$ resonances are simulated with a fixed $v_2$ of 6\% as in ~\cite{Feng:2018chm}, and all resonances have same $p_{T}$ for which the value itself can vary between simulations. Primordial pions are simulated again with realistic flow and spectra as aforementioned.  
In Fig.~\ref{fig:resFixedPtChange_typicalPrimFlowResFixed6PctV2NoResV3_compareA1} one can see that the $p_{T}$ change, over a range of $0.5 - 2$ GeV/$c$, has a visible effect on the observables. It is worth mentioning that, although a dedicated study has been devoted to this effect, it is not an additional, independent effect on top of existing effects already presented in the paper. This effect has been taken into consideration automatically when taking a characteristic $v_{2}$ and transverse spectra in simulations. However, it would be an interesting study in terms of understanding how the choice of slope (which changes $\langle p_{T} \rangle$) of transverse spectra would affect our observables. To investigate this, in Fig.~\ref{fig:resTChange_typicalPrimResFlow_compareA1} primordial pions and $\rho$ resonances are simulated according to their corresponding default characteristic flow and spectra as mentioned earlier, and $r_{\mathrm{rest}}$ and $R_{B}$ are calculated for a series of temperature of $\rho$-resonance spectra around its nominal value of 317 MeV. The study is repeated for various $a_{1}$ values. One can find that $r_{\mathrm{rest}}$ changes for merely $\sim2\%$ relatively over a temperature span of 40\% change.

\section{AMPT and AVFD models}\label{sec:AMPT_AVFD}

In this section the two observables are examined with two popular realistic models, namely the AMPT and the AVFD models. 

 The AMPT model~\cite{Lin:2004en} uses the Heavy Ion Jet Interaction Generator (HIJING~\cite{Wang:1991hta,Wang:1991us}) for generating the initial conditions, the Zhang's Parton Cascade (ZPC~\cite{Zhang:1997ej}) for modeling the partonic scatterings, and A Relativistic Transport (ART~\cite{Li:1995pra,Li:2001xh}) model for treating hadronic scatterings.
 The version (v2.25t4cu) we used is a version with string melting, in which it treats the initial condition as partons and uses a simple coalescence model to describe hadronization. It is also a version with charge-conservation being assured~\cite{Lin:2019pri}, which is particularly important for the CME related model-studies.

The AVFD framework~\cite{Jiang:2016wve,Shi:2017cpu} implements the anomalous transport current from the CME into fluid dynamics framework to simulate the evolution of fermion currents on an event-by-event basis and to evaluate the resulting charge separation in QGP, on top of the neutral bulk background described by the VISH2+1 hydrodynamic simulations ~\cite{Song:2010mg} with Monte-Carlo Glauber initial conditions, followed by a URQMD hadron cascade stage ~\cite{Bleicher:1999xi,Bass:1998ca}. This new tool allows one to quantitatively and systematically investigate the CME signal and account for the resonance contributions. The version used in this paper is beta-1.0, with the level of local charge conservation set to be 33\%.

 Both AMPT and AVFD models are known to have a good description of experimental data, including particle's yield, spectra and flow. They can serve as good baselines for apparent charge separation arising from pure backgrounds. In addition, the CME feature implemented in AVFD will allow one to study the observable's response to signal in a relatively realistic environment of backgrounds. 

\begin{figure}[htbp]
\centering
\makebox[1cm]{\includegraphics[width=0.45 \textwidth]{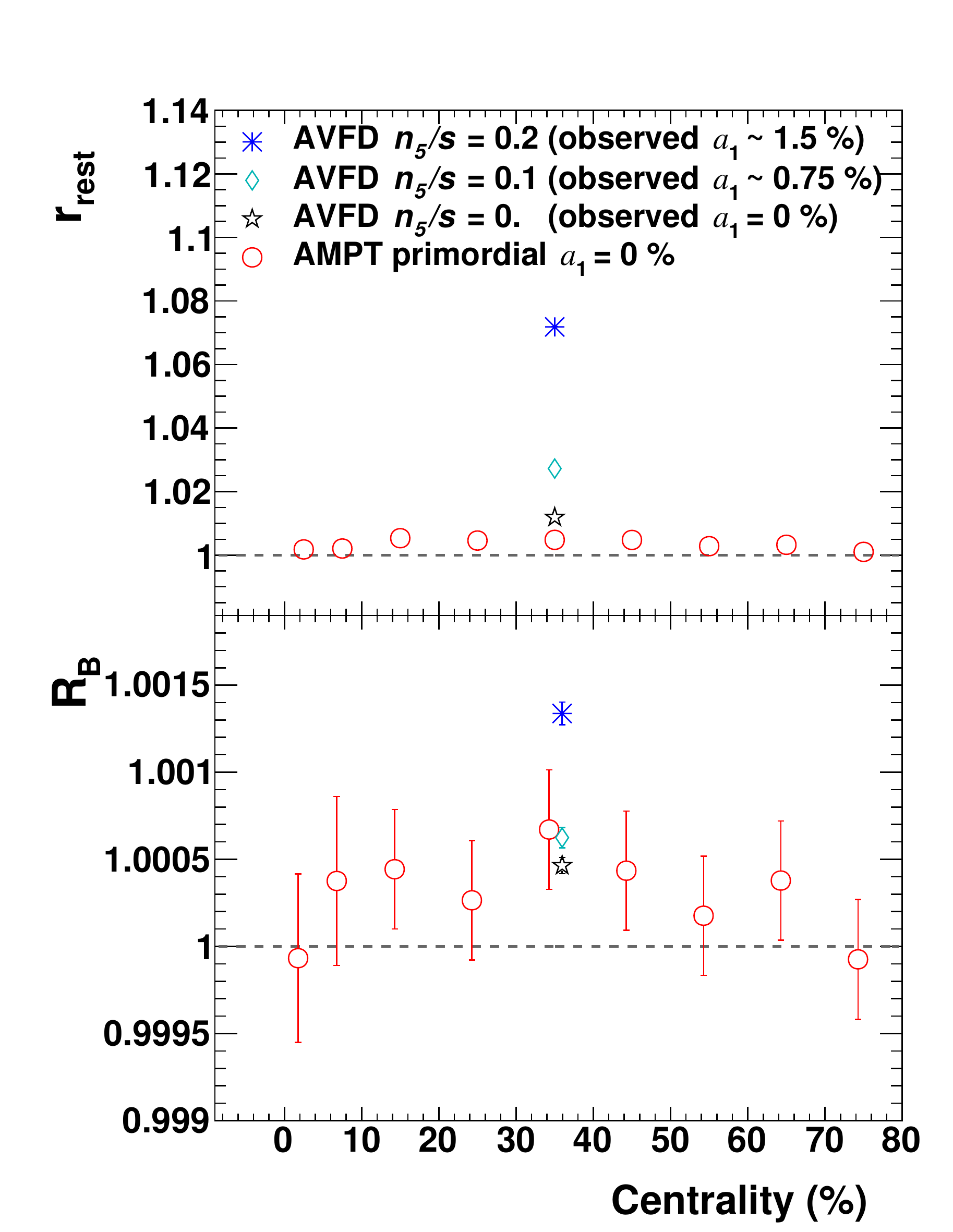}}
\caption{(Color online) $r_{\mathrm{rest}}$ and $R_{B}$ as a function of centrality, calculated for events from AMPT and AVFD models. The AMPT model has no built-in CME effect.  In the AVFD model the CME is implemented by finite ratio of axial charge over entropy ($n_{5}/s$), resulting in finite average $a_{1}$ (observed $a_{1}$) for all charged particles, including primordial ones and those from resonance decays. The LCC level is set to be 33\% in AVFD events.}
\label{fig:AMPT_AVFD}
\end{figure}

Fig.~\ref{fig:AMPT_AVFD} shows $r_{\mathrm{rest}}$ and $R_{B}$ as a function of centrality for AMPT and AVFD events. For AMPT, each point in the figure is calculated with $\sim2$ million model-events, and for AVFD, $\sim50$ million for cases with the CME (finite $n_5/s$), and $\sim 100$ million for the case without it ($n_5/s = 0$). To match typical acceptance cuts used by the STAR collaboration, only particles that satisfy $|\eta|<1$ and $0.2 < p_{T} < 2$ GeV/$c$ are considered in the analysis.
For the two cases without the CME (AMPT, and AVFD with $n_{5}/s = 0$), $r_{\mathrm{rest}}$ values is in between 1 and 1.015 depending on centrality, and is smallest if compared to cases with the CME. The AMPT study shows that $r_{\mathrm{rest}}$ in middle central collisions is in general larger than central and peripheral collisions, likely due to a convolution of the multiplicity effect and the apparent charge separation arising from backgrounds. $r_{\mathrm{rest}}$ increases clearly with increasing $n_{5}/s$, indicating a very good sensitivity to the CME.  A good sensitivity to the CME is also seen for $R_{B}$. In general the proposed observables behave as expected for realistic models. 

Note that realistic models like AMPT and AVFD include additional backgrounds from Transverse Momentum Conservation (TMC) and Local Charge Conservation (LCC), which may also introduce correlations to the CME observables, albeit the effect is not expected to be as strong as resonance flow. In Fig.~\ref{fig:AMPT_AVFD}, the small but above-unity values of $r_{rest}$ and $R_{B}$ for AMPT events (and AVFD events with $n_5/s=0$ as well) are indications of their existence. Such effects cannot be conveniently studied by toy models, they have to be addressed with realistic models like AMPT and AVFD.

\section{Summary}

In this article a pair of observables, $r_{\mathrm{rest}}$ and $R_{B}$, are presented as alternative ways to study the charge separation induced by the CME in relativistic heavy ion collisions.  Both observables have been studied with toy model simulations, as well as two realistic models, namely, AMPT and AVFD. The toy model studies include flow-related backgrounds, and for the first time, backgrounds that are related to the global spin alignment of resonances. It is shown that the two observables have similar positive responses to signal, and opposite, limited responses to identifiable backgrounds arising from resonance flow and global spin alignment. This information can be useful under certain scenarios in identifying charge separation induced by backgrounds. For example, if both $r_{\mathrm{rest}}$ and $R_{B}$ are above unity, then one has a case in favor of the existence of the CME. 

However, even with both $r_{\mathrm{rest}}$ and $R_{B}$ being greater than unity, there are remaining backgrounds arising from momentum and charge conservation which have to be studied in details with realistic models. Like any other approach, this procedure does not provide a complete, clean solution under all possible scenarios. A quantitative statement on signal versus background has to rely on realistic simulations, and, better to be made with the help of additional, external information (such as information from isobaric collisions). That said, the two observables do provide useful insights into the problem from a unique perspective.

\section*{Acknowledgements}

 The author is grateful to G. Wang, J. Liao, S. Shi and N. Magdy for fruitful discussions. In particular the author thanks G. Wang for stimulating discussions that lead to the initiation of this study, as well as a conversation at a later time that facilitates the explanation of the effect of global spin alignment. Additional thanks go to Z. Lin, G. Ma and G. Wang for providing AMPT events, and S. Shi and J. Liao for providing AVFD events. The author also thanks H. Ke for his help in allocating computing resources, and G. Wang and Y. Lin for their help in facilitating the processing of AVFD events. The author thanks G. Wang, J. Liao and Z. Lin for reading the manuscript and providing comments. A.H. Tang is supported by the US Department of Energy under Grants No. DE-AC02-98CH10886 and No. DE-FG02-89ER40531. 

{}
\end{document}